\documentclass[review]{elsarticle}

\usepackage{lineno,hyperref}
\modulolinenumbers[5]

\usepackage{lineno}
\usepackage{latexsym,amssymb,amsmath}
\usepackage{mathabx}
\usepackage{longtable}
\usepackage{tikz,pgf}
\usepackage{subfig}
\usepackage{wrapfig}
\usepackage{rotating}
\usepackage{multirow}
\usepackage{hyperref}
\usepackage{umoline}
\usepackage{enumitem}
\usepackage{booktabs}

\usepackage{amsthm}
\theoremstyle{plain}
\newtheorem{theorem}{Theorem}[section]

\theoremstyle{definition}

\newtheorem{example}{Example}[section]

\theoremstyle{remark}

\usepackage{algorithm}
\usepackage{algorithmic}

\usepackage{graphicx}
\usepackage{epstopdf}
\input{def.set}

\usepackage{url}
\usepackage{hyperref}
\hypersetup{colorlinks,%
citecolor=blue,%
filecolor=blue,%
linkcolor=red,%
urlcolor=blue,%
pdftex}

\journal{Journal of \LaTeX\ Templates}

\begin{document}

\begin{frontmatter}

\title{Volume entropy and generalized Markov system \\ for modelling information flow in a brain graph}

\author[a,b]{Hyekyoung Lee}
\author[a,c]{Eunkyung Kim}
\author[b,d]{Hyejin Kang}
\author[b,e]{Youngmin Huh} 
\author[d,f]{Youngjo Lee} 
\author[h]{Seonhee Lim\corref{mycorrespondingauthor1}}
\cortext[mycorrespondingauthor1]{Corresponding author}
\ead{slim@snu.ac.kr}
\author[b,e,g,i]{Dong Soo Lee\corref{mycorrespondingauthor2}}
\cortext[mycorrespondingauthor2]{Corresponding author}
\ead{dsl@plaza.snu.ac.kr}

\address[a]{Biomedical Research Institute, Seoul National University Hospital} 
\address[b]{Department of Nuclear Medicine, }
\address[c]{Department of Rehabilitation Medicine, Seoul National University College of Medicine,}
\address[d]{Data Science for Knowledge Creation Research Center,}
\address[e]{Interdisciplinary Program in Cognitive Science,} 
\address[f]{Department of Statistics, College of Natural Sciences,}
\address[g]{Department of Molecular Medicine and Biopharmaceutical Sciences, Graduate School of Convergence Science and Technology, and College of Medicine or College of Pharmacy, 
Seoul National University}
\address[h]{Department of Mathematical Sciences, Seoul National University, Seoul}
\address[i]{Korea Brain Research Institute, Daegu, Republic of Korea}
%

%
%
%

\begin{abstract} 
Entropy is a classical measure to quantify the amount of information or complexity of a system. 
Various entropy-based measures such as functional and spectral entropies have been proposed in brain network analysis. 
However, they are less widely used than traditional graph theoretic measures such as global and local efficiencies because either they are not well-defined on a graph or difficult to interpret its biological meaning. 
In this paper, we propose a new entropy-based graph invariant, called volume entropy. 
It measures the exponential growth rate of the number of paths in a graph, which is a relevant measure if information flows through the graph forever. 
We model the information propagation on a graph by the generalized Markov system associated to the weighted edge-transition matrix. 
We estimate the volume entropy using the stationary equation of the generalized Markov system. 
A prominent advantage of using the stationary equation is that it assigns certain distribution of weights on the edges of the brain graph, which we call the stationary distribution. 
The stationary distribution shows the information capacity of edges and the direction of information flow on a brain graph.
The simulation results show that the volume entropy distinguishes the underlying graph topology and geometry better than the existing graph measures. 
In brain imaging data application, the volume entropy of brain graphs was significantly related to healthy normal aging from 20s to 60s. 
In addition, the stationary distribution of information propagation gives a new insight into the information flow of functional brain graph. 
\end{abstract} 

\end{frontmatter}

\section{Introduction}
\label{sec:intro} 

Brain regions not only function individually but also are functionally linked to each other. 
We consider the brain as a network whose nodes are brain regions, which are connected to each other according to the intensity of their functional links \cite{sporns.2010.book}. 
The functional  connection  between  brain regions is anti-proportional to the interregional correlation of brain regions on brain imaging studies such as functional magnetic resonance imaging (fMRI) or positron emission tomography  (PET) \cite{brier.2013.na,sanz-argita.2010.plosone,toussaint.2012.ni,vanheuvel.2008.ni,wang.2007.hbm,zhou.2006.phrl}. 
On efficient brain network, the information is quickly transmitted over the whole brain because there are sufficient short paths between all the pairs of brain regions.  
Furthermore, when we say brain networks are locally efficient, it implies that sufficient alternative paths exist between brain regions, even when a connection is damaged and nonfunctioning. 
These global and local efficiencies of brain networks change along with aging \cite{daianu.2013.bc,ferreira.2013.nbr,greicius.2004.pnas}. 
The change of a network has been quantified by complex graph theoretic measures including global and local efficiencies, characteristic path length, and clustering coefficient \cite{bullmore.2012.nrn,vanheuvel.2009.jneuro,rubinov.2009.ni}.  
They were proposed to be used as network-based biomarkers of normal aging.

Information entropy measures the average amount of information of a system in information theory \cite{cover.2006.book}.
It is clearly defined as the expected value of the negative logarithm of the probability distribution of the system. 
Various kinds of information entropies, especially functional and spectral entropies, have been proposed as a graph invariant of a human brain \cite{yao.2013.sr,sato.2013.ni}.  
Functional entropy is the information entropy which uses the relative frequency histogram of edge weights for the probability distribution \cite{yao.2013.sr}. 
Therefore, functional entropy cannot distinguish two networks with different topologies and with the same histogram of edge weights. 
Spectral entropy is the information entropy defined with the eigenvalues of the adjacency matrix of an unweighted network \cite{sato.2013.ni}. 
The eigenvalues of the adjacency matrix provide some information about the shape of a graph, especially that related to connectedness. 
However, the relationship between the eigenvalues and the connectedness is not clearly established. 
Thus, it is difficult to interpret its biological meaning. 
Like the functional entropy, the spectral entropy also needs the procedure to approximate the probability distribution of the eigenvalues using Gaussian kernel regression.  
The approximation procedure includes the parameter selection such as the number of bins in the functional entropy and the width of Gaussian in the spectral entropy.  

Topological entropy is the complexity measure of a topological dynamical system in ergodic theory and geometry \cite{frigg.2011.pp}. 
It measures how much energy has flowed in the system or how widely it has spread out over the system \cite{frigg.2011.pp}. 
The topological entropy of the geodesic flow on a graph is called volume entropy \cite{lim.2008.tams}. 
The volume entropy assumes that the information flows through the links on a brain graph.  
If the time goes to infinity, the network paths through which information flow will increase  exponentially.   
The volume entropy is the exponential growth rate of the number of network paths through which information flows. 
The larger the volume entropy is, the more information flows on the graph. 
In this study, we introduce the volume entropy as a new invariant of brain networks. 
To compute the volume entropy, we model the information flow on a graph by the generalized Markov system associated to a new edge-transition matrix \cite{lim.2008.tams}. 
The volume entropy is obtained by the stationary equation of the generalized Markov system. 
Furthermore, its stationary distribution shows the edge capacity of information as well as the direction of information flow on a network. 
Thus, we can derive a directed network that represents the information propagation on a graph at the stationary state. 

In simulations, we compared the volume entropy of various artificial graphs such as regular, small-world, random, scale-free, hyperbolic, and modular graphs. 
The results showed that the volume entropy distinguished the underlying graph topology and geometry better than the existing network measures such as global and local efficiencies and entropy-based invariants. 
We also applied the volume entropy to the resting state fMRI and PET data obtained from 38 normal controls between the ages of 20s and 60s. 
The volume entropy revealed the change of information flows in a brain graph during healthy normal aging. 
The main contributions of our work were as follows:  
\begin{itemize}[noitemsep,topsep=0pt]
\item We introduced the new invariant of brain graphs, called volume entropy. It was supposed to quantify the efficiency of a brain graph in terms of information propagation.  
\item The information flow on a brain graph was modelled by the generalized Markov system associated with a newly defined edge-transition matrix.  We also derived information flow on a brain graph based on the stationary equation of the generalized Markov system. 
\item The proposed method was applied to the functional and metabolic graphs obtained from resting state fMRI and PET data, respectively. The results revealed the information propagation on a brain graph changing with the age.     
\end{itemize}

\section{Materials and methods} 
\label{sec:methods}

\subsection{Resting state fMRI and PET data sets} 
\label{sec:data_set}

PET and fMRI data were simultaneously acquired from 38 healthy normal subjects (M/F: 19/18, age: $43.9\pm13.9$) from 20s to 60s using a Siemens Biograph mMR 3T scanner (Siemens Healthcare Sector, Germany). 
MR images had 116 volume of images per a subject. The first 4 volumes were discarded among 116 volumes and 112 volume of images a subject were used for network analysis. 
After preprocessing using the AFNI \cite{cox.1996.cbr} and the FSL \cite{smith.2004.ni}, we parcellated the brain into 116 regions of interest (ROIs) according to automated anatomical labelling (AAL) \cite{tzourio-mazoyer.2002.ni}. 
Among the 116 regions, 90 brain regions were selected as the nodes and 26 cerebellar regions were not included in a graph (the number of nodes, $p=90$). 
The measurement of each node was obtained by averaging blood-oxygen-level dependent (BOLD) signals in the ROI of fMRI data. 
Each node had $n$ measurements, which were the number of time points a subject in fMRI data ($n=112$). 
The measurement vectors of 90 ROIs were written by $\bx_{1}^{j}, \dots \bx_{p}^{j} \in \Real^{n}$ of the $j$th subject ($j=1,\dots,38$, $p=90$, $n=112$).   

PET images were preprocessed using the Statistical Parametric Mapping (SPM8, www.fil.ion.ucl.ac.uk/spm) and PVElab software \cite{quarantelli.2004.jnm}. 
The image intensity of gray matter was globally normalized to 50. 
The measurement of a node was obtained by averaging FDG uptakes in the corresponding ROI. 
We divided the data into two groups, young (age: $32.2\pm6.9$) and old (age: $55.6\pm7.7$) depending on whether a subject was over age 45. 
The number of subjects in each group was 19. 
We had the measurement vectors of PET data, $\bx_{1}^{Y}, \dots, \bx_{p}^{Y} \in \Real^{n}$ for young group, denoted by Y, and $\bx_{1}^{O}, \dots, \bx_{p}^{O} \in \Real^{n}$ for old group, denoted by O ($p=90$, $n=19$). 

\subsection{Distance of brain network}
\label{sec:brain_network_construction}

The edge weight between two nodes $i$ and $t$ is estimated by the Gaussian kernel based on Pearson correlation: 
\be
\label{eq:gaussian_kernel} 
w_{it} = k(\bx_{i},\bx_{t}) = \exp{\left(-\frac{1-corr(\bx_{i},\bx_{t})}{\sigma_{i}\sigma_{t}}\right)}, 
\ee
where $corr(\bx_{i},\bx_{t})$ is the Pearson correlation between two measurement vectors $\bx_{i}$ and $\bx_{t}$ and $\sigma_{i}$ is the width of Gaussian kernel. 
Because $1 - corr(\bx_{i},\bx_{t}) = \frac{1}{2} \parallel \frac{\bx_{i}}{\parallel \bx_{i} \parallel} - \frac{\bx_{t}}{\parallel \bx_{t} \parallel} \parallel^{2}$, $\sqrt{1-corr(\bx_{i},\bx_{t})}$ is conditionally negative semi-definite for $\bx_{i},\bx_{t} \in \Real^{n}$ \cite{jayasumana.2015.pami}. 
The Gaussian kernel based on correlation in (\ref{eq:gaussian_kernel}) is positive definite for all $\sigma_{i} > 0$ and satisfies Mercer's theorem \cite{jayasumana.2015.pami}. 
Thus, it transforms the original data in a nonlinear manifold into a higher dimensional feature space where the transformed features have a linear representation.  
The distance of the kernel $w_{it}$ is estimated by a kernel trick \cite{scholkopf.2001.nips}: 
\be
\label{eq:kernel_distance}
d_{it} &=& \parallel \phi(\bx_{i}) - \phi(\bx_{t}) \parallel \nonumber \\ 
&=& \left< \phi(\bx_{i}) - \phi(\bx_{t}),\phi(\bx_{i}) - \phi(\bx_{t}) \right>^{1/2} \nonumber \\ 
&=& [k(\bx_{i},\bx_{i}) + k(\bx_{t},\bx_{t}) - 2 k(\bx_{i},\bx_{t})]^{1/2} \nonumber \\ 
&=& \sqrt{2 - 2 k(\bx_{i},\bx_{t})}. 
\ee
If an edge $e$ connects two nodes $i$ and $t$, $d_{it}$ is also denoted by $l(e)$. 

The kernel-based distance is a Euclidean distance between two nodes in a higher dimensional feature space. 
When the kernel width is small in (\ref{eq:gaussian_kernel}), the local neighbors that are highly positively correlated in the original data space are more clearly separated in the feature space, while non-local neighbors are not.  
The kernel width $\sigma_{i}$ in (\ref{eq:gaussian_kernel}) is determined by the tenth smallest one among all $1-corr(\bx_{i}, \bx_{t})$ $(t=1,\dots,i-1,i+1,p)$ \cite{zelnik-manor.2004.nips}.  

The 38 brain graphs of 38 subjects were constructed from fMRI data by the kernel-based distance in (\ref{eq:kernel_distance}) and (\ref{eq:gaussian_kernel}). 
Two brain graphs of two groups, Y and O were constructed from PET data. 
We call the brain graphs constructed by fMRI and PET data functional and metabolic graphs, respectively.   

\subsection{Volume entropy} 

Suppose that $\calN = \calN(V,E)$ is a connected finite graph with the node set $V$ and the edge set $E$. 
We will assume that $\calN$ does not have any terminal node. We will be given a length $l(e)$ for each edge $e$, which determines a distance on $\calN$.
Let $S$ be a subset of edges with multiplicities, where each edge can be counted several times. 
The length of $S$ is defined by $l(S) = \sum_{e \in S} l(e)$. 
For example, $S = \left\{ e,e,f \right\}$ for $e,f \in E$ is allowed and the volume of $S$ is $2l(e) + l(f)$. 
Edge $e$ is assumed to have an orientation from the initial node $i(e) \in V$ to the terminal node $t(e) \in V$. For a given oriented edge $e$ from $i(e)$ to $t(e)$, we denote by $\overline{e}$ the oriented edge from $i(\overline{e})=t(e)$ to $t(\overline{e})=i(e)$.
Note that for any $e \in E$, both $e$ and $\overline{e}$ exist in $\calN$. 
  
The sequence of $n$ consecutive edges without backtracking is denoted by a path $\mathcal{P} = e_{1} e_{2}  \cdots e_{n}$ ($e_{j+1}  \neq \overline{e_j}$, $e_{j} \in E$). 
The set of all possible paths of length $r$ starting from a node $v_{0} \in V$ in $\calN$ has a structure of a tree, which we denote by $B(v_{0},r)$. 
Because $\calN$ is assumed to have no terminal node, the number of possible paths $B(v_0, r)$ increases exponentially as $r \rightarrow \infty$. 
The limit of the ball $B(v_0,r)$ as $r \to \infty$ is called the universal covering tree of $\calN$. 

\begin{figure*}[t] \begin{center}
\includegraphics[width=1\linewidth]{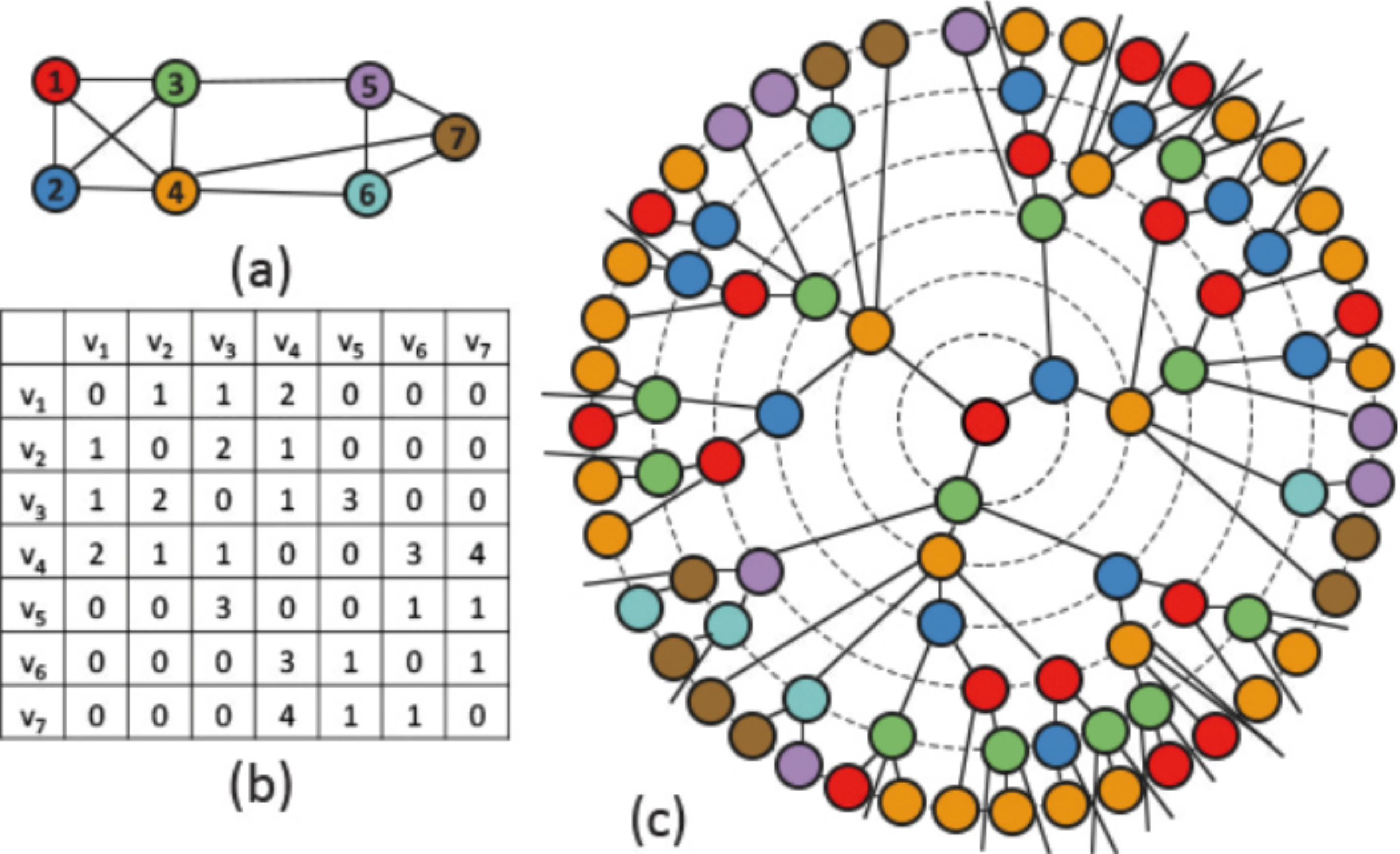} \caption{(a) Toy example of a weighted graph that consists of 7 nodes and 12 edges. The degree of all nodes is more than three. (b) Distance matrix of (a). (c) Tree $B(v_{0},r)$ starting from $v_{0}=v_{1}$ with the radius $r=6$. All nodes in (c) have the same connections to nodes in (a).} \label{fig:universal_covering_tree} \end{center} \end{figure*}

\begin{example} 
\label{ex:toy_network} 
Fig. \ref{fig:universal_covering_tree} shows the universal covering tree of a toy example. 
The weighted graph and its distance matrix are given in (a) and (b), respectively. 
The tree $B(v_{1},6)$ of (a) is illustrated in (c). 
Each node in the tree (c) has the same edges to the neighboring nodes as the corresponding node in the graph (a). 
\end{example} 

The volume entropy $h_{vol}$ is defined as \cite{lim.2008.tams} 
\be
\label{eq:hvol} 
h_{vol} = \lim_{r \rightarrow \infty} \frac{\log l(B(v_{0},r))}{r}.
\ee  
The volume entropy $h_{vol}$ does not depend on $v_{0}$. When $h_{vol}>0$, it is easy to see that $l(B(v_0, r))$ is concentrated on the outer shell, i.e. $$l(B(v_0, r) - B(v_0, r-r_0)) \sim e^{h_{vol} r} - e^{h_{vol}(r-r_0)} = e^{h_{vol} r} (1-e^{-r_0}).$$ It follows that
$$ h_{vol}= \lim_{r \rightarrow \infty} \frac{\log l(B(v_{0},r) - B(v_{0},r-r_0))}{r} , \mbox{ for any } r_0>0.$$ 
Note also that $$ h_{vol} = \lim_{r \rightarrow \infty} \frac{\log N_{r} (v_{0})}{r},$$ where $N_{r} (v_{0})$ is the number of paths of length $r$ in the graph starting from $v_{0}$,    
since $$r_0 N_{r-r_0} (v_{0}) \le l(B(v_{0},r) - B(v_{0},r-r_0)) \le r_0 N_{r} (v_{0}).$$
In other words, the volume entropy $h_{vol}$ is the exponential growth rate of the number of paths $N_{r} (v_{0})$ as $r \to \infty$. 

\subsection{Generalized Markov system} 
\label{sec:calculation}

In this section, we will model the growth of the number of paths in a graph as $r \to \infty$ \cite{lim.2008.tams}. 
Recall that in any given graph $\calN (V,E)$, for every edge $e$, the inverse edge $\overline{e}$ is also in $\calN$ and $l(e) = l(\overline{e})$.
Denote by $q$ the number of oriented edges in $E$ counted with multiplicity. 

An \emph{edge-transition matrix} $\bL (h) = [L_{ef} (h)] \in \Real^{q \times q}$ is defined by 
$$L_{ef} (h) = a_{ef} e^{-h l(f)},$$
where $l(f)$ is the distance of an edge $f$ and $$\mbox{ } a_{ef} = \left\{ \begin{array}{cl} 1 & \mbox{if } t(e) = i(f), i(e) \neq t(f) \\ 0 & \mbox{otherwise} \end{array} \right. .$$
The \emph{generalized Markov system} of $\calN$ associated to $\bL(h)$ is defined by 
\be
\label{eq:eig}
\bz_{t} = \bL (h) \bz_{t+1},  
\ee
for $h > 0, t > 0,$ and $\bz_{t} \in \Real^{q}$. 
The dimension of $\bz_{t}$ is equal to the number of oriented edges in the graph. 
The entry of $\bz_{t}$ is proportional to the number of paths in the graph that go through the corresponding edge at time $t$.   
When $\bz = \bz_{t+1} = \bz_{t}$, (\ref{eq:eig}) is called the stationary equation of the generalized Markov system.  

\begin{theorem}[Theorem 4 in \cite{lim.2008.tams}]
Given a graph $\calN = \calN (V,E)$, the volume entropy of $\calN$ is $h_{vol}=h$ such 
that the generalized Markov system in (\ref{eq:eig}) is stationary, i.e., 
\be
\label{eq:eig_stationary} 
\bz = \bL(h) \bz.
\ee 
\end{theorem} 
$\bz$ in (\ref{eq:eig_stationary}) is the eigenvector of $\bL(h_{vol})$ with the largest eigenvalue $1$.  
If we denote the entry of $\bz$ as $z_{e}$ in (\ref{eq:eig}), the sum of the squares of all $z_{e}$s is equal to one, i.e., $\sum_{e} z_{e}^2=1$. 
Moreover, $\bz$ has all entries of the same sign according to the Perron-Frobenius theorem.  
Thus, we call $\bz.^2=[z_{e}^2]$ the stationary distribution of the generalized Markov system associated to the edge-transition matrix $\bL (h)$. 
\begin{wrapfigure}{r}{0.4\textwidth} \begin{center}
\includegraphics[width=1\linewidth]{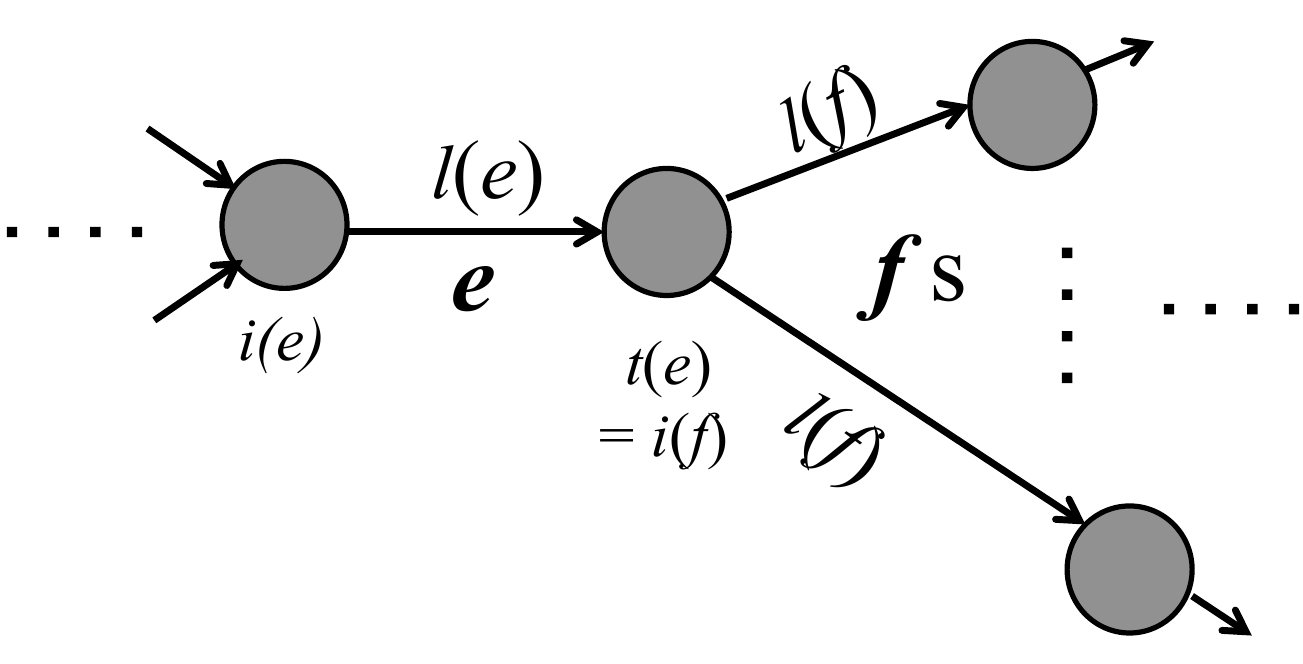} \caption{Paths from $e$ to $f$s} \label{fig:eq6} \end{center} \end{wrapfigure}
The generalized Markov system in (\ref{eq:eig_stationary}) can be rewritten in the scalar form by 
\be
\label{eq:entry_eig}
z_{e} = \sum_{f} a_{ef} e^{-h_{vol} l(f)} z_{f}. 
\ee
The equation (\ref{eq:entry_eig}) implies that the number of possible paths increases exponentially with the growth rate $h_{vol} l(f)$.  

We reshape the vector $\bz.^2$ into a matrix, $\bPi = [\pi_{it} = z_{e}^{2}] \in \Real^{p \times p},$ where an edge $e$ has the initial node $i(e) = i$ and the terminal node $t(e) = t$ and $\pi_{ii}=0$  $(i = 1, \dots, p)$. 
$\pi_{it} (= z_{e}^{2})$ is related to the number of paths in the graph that go through the edge $e$ at the stationary state.  
Thus, we call $\pi_{it} (= z_{e}^{2})$ an edge capacity and $\bPi$ an edge capacity matrix. 
Since $\bPi$ is an asymmetric matrix, $\bPi$ is a directed graph where the node and edge sets were $V$ and $E$ of the given graph $\calN$, and the edge weight was the edge capacity. 
In the stationary equation (\ref{eq:entry_eig}), $z_{e}(=\sqrt{\pi_{it}})$ is affected by the distance $l(f)$ for all edges $f$s connected with the terminal node $t$. 
Thus, $\pi_{it}$ is different from $\pi_{ti}$. The difference between $\pi_{it}$ and $\pi_{ti}$ is related to the imbalance of the connectivities of two nodes $i$ and $t$. 
We define a node capacity by the difference between the inward and outward edge capacities of a node, estimated by $\pi_{i} = \sum_{t} \pi_{ti} - \sum_{t} \pi_{it}$.  
If a node capacity is negative/positive, the outgoing edge capacities larger/smaller than the incoming edge capacities.

\begin{figure*}[t] \begin{center}
\includegraphics[width=1\linewidth]{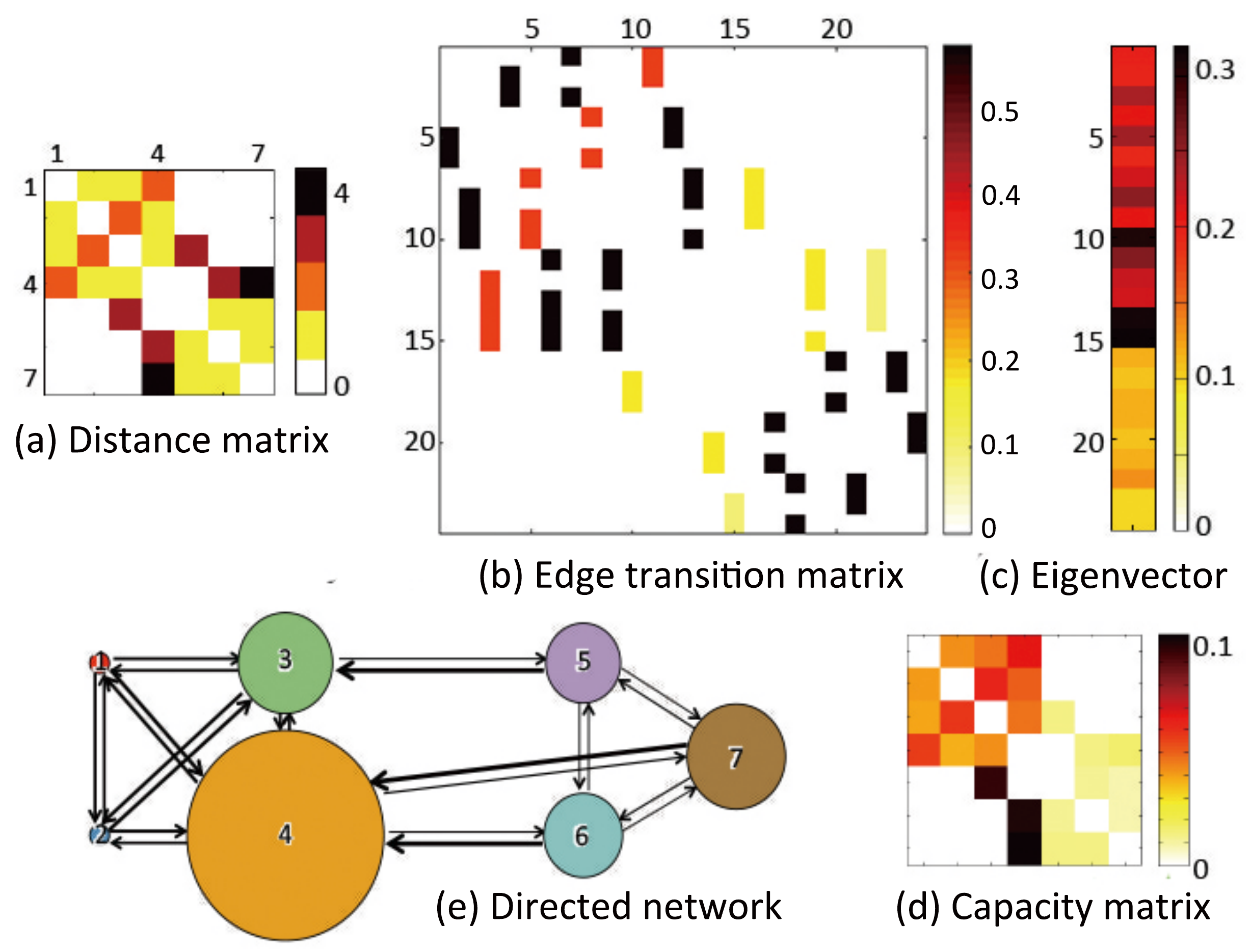} \caption{(a) Distance matrix of the graph in Fig. \ref{fig:universal_covering_tree}. (b) Edge-transition matrix $\bL(h_{vol}) \in \Real^{24 \times 24}$. If edges $e$ and $f$ are consecutive, the $(e,f)$th entry is $e^{-h_{vol}l(f)}$. (c) Eigenvector $\bz=[z_{e}]\in\Real^{24}$. (e) Edge capacity matrix $\bPi = [\pi_{ij} = z_{e}^2]$, where an edge $e$ connects between a node $i$ and $j$. (d) Directed graph induced by $\bPi$. The line width is proportional to the entry of $\bPi=[\pi_{ij}]$. The size of the node $i$ is proportional to the absolute of node capacity, $|\pi_{i} = \sum_{j} \pi_{ji} - \sum_{j} \pi_{ij}|$.  } \label{fig:toy_network_path} \end{center} \end{figure*}

\begin{example} 
\label{ex:toy_directed} 
Given the weighted graph in the toy example \ref{ex:toy_network}, Fig. \ref{fig:toy_network_path} shows (b) its edge-transition matrix $\bL(h_{vol})$, (c) eigenvector $\bz$, (d) edge capacity matrix $\bPi$, and (e) the induced directed network. 
The example of the graph in Fig. \ref{fig:universal_covering_tree} (a) had $7$ nodes and $12$ edges. 
To estimate volume entropy, each undirected edge is assumed to consist of bidirectional edges with the same distance. 
The number of oriented edges is $q = 12 \cdot 2=24$. 
Thus, $\bL(h_{vol})$ is a $24\times24$ dimensional sparse matrix in Fig. \ref{fig:toy_network_path} (b), and its eigenvector $\bz$ is a $24-$dimensional vector in (c).    
Fig. \ref{fig:toy_network_path} (e) shows the directed network induced by the edge capacity matrix $\bPi$ in (d). 
The line width of edge and size of node are proportional to the edge and node capacities, respectively. 
In the original graph in Fig. \ref{fig:universal_covering_tree} (a) , the node sets $\left\{ 1, 2, 3,4 \right\}$ and $\left\{ 5,6,7 \right\}$ form a clique, respectively called A and B for convenience. 
If we defined the module size by the number of nodes, the module size of A is larger than that of B. 
Thus, there are more paths for information flow within A than B. 
More paths were directed from B to A   
through three edges from the node 5 to 3, from 7 to 4, and from 6 to 4. 
In the induced directed network in Fig. \ref{fig:toy_network_path} (e), while the bidirectional edges within A or B have similar line width (edge capacity), the width of edges from A to B are much different from that from B to A.  
\end{example} 

\subsection{Normalization of graph volume} 
\label{sec:normalization}

The most of graph invariants including the volume entropy are influenced by the volume of graph. 
If we denote the sum of all edges in $\calN(V,E)$ as $vol(E)$ and the unnormalized edge distance of $f$ as $\tilde{l}(f)$, the normalized distance $l(f)$ in (\ref{eq:entry_eig}) is obtained by $l(f) = \frac{2}{vol(E)} \tilde{l}(f)$. 
The stationary equation (\ref{eq:entry_eig}) is rewritten by 
$$z_{e} = \sum_{f} a_{ef} e^{- \frac{2}{vol(E)} h_{vol} \tilde{l}(f)} z_{f}.$$ 
Then, the volume entropy of unnormalized graph is obtained by  
$$ \tilde{h}_{vol} = \frac{2}{vol(E)} h_{vol}.$$
The stationary distribution $\bz.^2$ does not depend on the normalization of graph. 
In this study, we estimated the normalized volume entropy for all weighted graphs. 

%

\section{Comparisons of network invariants using three simulated data sets} 

\subsection{Using artificial unweighted graphs}  
\label{sec:toy_binary}


In this simulation, we compared the performance of six global graph invariants in distinguishing five artificial unweighted graphs with varying the sparsity of a graph.  
The six different global graph invariants were as follows: 
\begin{itemize}[noitemsep,topsep=0pt]
\item global efficiency $e_{glo}$,  
\item average local efficiency $e_{loc}$, 
\item modularity $Q$,     
\item functional entropy $h_{fun}$ \cite{yao.2013.sr},  
\item spectral entropy $h_{spe}$ \cite{sato.2013.ni}, and  
\item volume entropy $h_{vol}$ (the proposed method), 
\end{itemize}
The five artificial unweighted graphs that were used for the comparison of performance were as follows: 
\begin{itemize}[noitemsep,topsep=0pt]
\item regular graph (RE),   
\item small-world graph (SW),   
\item random graph (RA),   
\item scale-free graph (SF),  and 
\item hyperbolic graph (HY).    
\end{itemize} 
RE is an unweighted graph where all nodes have the same degree. 
SW is a globally and locally efficient graph with short characteristic path length and large average clustering coefficient \cite{bassett.2006.neuroscientist}.
SF has heterogeneous degree distribution with a few number of heavily linked nodes, termed hubs, but many nodes with few connections \cite{eguiluz.2005.phrl}. 
Hubs make a great contribution to propagating information quickly throughout a network. 
On the other hand, it is vulnerable to targeted attacks on hubs. 
Thus, SF is known to be globally efficient and locally inefficient. 
HY is known to have both strong heterogeneity and high clustering coefficient \cite{krioukov.2010.pre}. 
It can be thought as a maximally efficient unweighted graph.
These
unweighted graphs were generated by CNM matlab toolbox \cite{alanis-lobato.2014.toolbox}. 

The number of nodes was fixed by $p=90$. 
The sparsity, which was the ratio of the number of edges to the number of maximally possible edges, was varied from 0.04 to 0.90.  
All nodes in a graph should have more than three edges for the estimation of volume entropy \cite{lim.2008.tams}. 
If there were nodes with degree less than three in the generated graph, we randomly took an edge connecting nodes with degree more than four and rewired it to a node with degree less than three.  
In this way, we generated 150 artificial unweighted graphs for each sparsity and each graph type. 
After five invariants were estimated in each graph, Wilcoxon rank sum test was performed to assess the statistical difference of each invariant between graph types at each sparsity. 
We used brain connectivity toolbox for the estimation of global and local efficiencies and modularity \cite{rubinov.2009.ni}. 

\begin{figure}[t] \begin{center}
\includegraphics[width=1\linewidth]{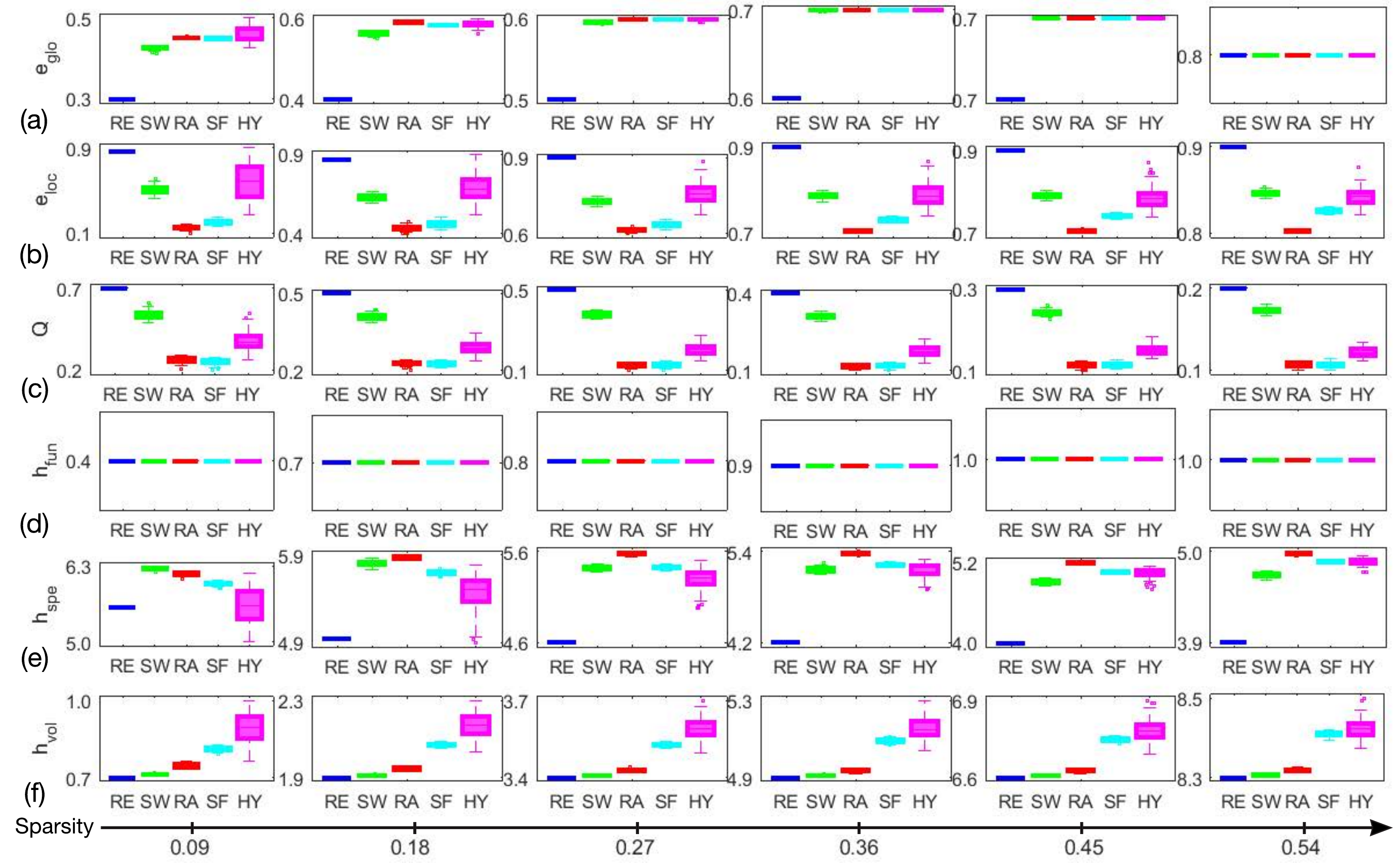} \caption{Comparison of graph invariants using various artificial unweighted graphs. Each panel shows the box plot of (a) blobal efficiency ($e_{glo}$), (b) average local efficiency ($e_{loc}$), (c) modularity ($Q$), (d) functional entropy ($h_{fun}$), (e)  spectral entropy ($h_{spe}$), and (f) volume entropy ($h_{vol}$). The sparsity of unweighted network was $0.09, 0.18, 0.27, 0.36, 0.45,$ and $0.54$ from left to right. The color of line represents RE (blue), SW (green), RA (red), SF (cyan), and magenta (HY). The order of the five types of different unweighted graphs was consistent for the sparsity when the graph property was measured by modularity and volume entropy in (c) and (f), respectively. However, the order was changed more than three times depending on sparsity in the other graph invariants, (a) $e_{glo}$, (b) $e_{loc}$, and (e) $h_{spe}$. $h_{fun}$ in (d) was exactly the same for all graph types at the fixed sparsity.  \\ (In (a) $e_{glo}$, the order of the five types of unweighted graphs was (1) $\mbox{RE}<\mbox{SW}<\mbox{RA}<\mbox{SF}<\mbox{HY}$ in the sparsity $[0.04 \mbox{ } 0.12)$, (2) $\mbox{RE}<\mbox{SW}<\mbox{SF}<\mbox{RA}<\mbox{HY}$  in $[0.12 \mbox{ }  0.20)$, (3) $\mbox{RE}<\mbox{SW}<\mbox{SF}\le\mbox{HY}<\mbox{RA}$  in $[0.20  \mbox{ } 0.48)$, and (4) $\mbox{RE}=\mbox{SW}=\mbox{RA}=\mbox{SF}=\mbox{HY}$  in $[0.48  \mbox{ } 0.90]$ ($p<.001$). In (b) $e_{loc}$, the order was (1) $\mbox{RA}<\mbox{SF}<\mbox{HY}<\mbox{SW}<\mbox{RE}$ in $[0.04  \mbox{ } 0.08)$, (2) $\mbox{RA}<\mbox{SF}<\mbox{SW}\le\mbox{HY}<\mbox{RE}$ in $[0.08  \mbox{ } 0.40)$, (3) $\mbox{RA}<\mbox{SF}<\mbox{HY}\le\mbox{SW}<\mbox{RA}$ in $[0.40  \mbox{ } 0.66)$, and (4) $\mbox{RA}<\mbox{SW}<\mbox{SF}<\mbox{HY}<\mbox{RE}$ in $[0.66  \mbox{ } 0.90]$ ($p<.001$). In (c) $Q$, the order  was $\mbox{RA} \le\mbox{SF}<\mbox{HY}<\mbox{SW}<\mbox{RE}$ for all sparsity. In (d) $h_{fun}$, the order was $\mbox{RE} = \mbox{SW}=\mbox{RA}=\mbox{SF}<\mbox{HY}$ for all sparisty. 
In (e) $h_{spe}$, the order was (1) $\mbox{HY}<\mbox{RE}<\mbox{SF}<\mbox{RA}<\mbox{SW}$  in $[0.04  \mbox{ } 0.12)$, (2) $\mbox{RE}<\mbox{HY}<\mbox{SF}<\mbox{RA}<\mbox{SW}$ in $[0.12  \mbox{ } 0.16)$, (3) $\mbox{RE}<\mbox{HY}<\mbox{SF}<\mbox{SW}<\mbox{RA}$ in $[0.16  \mbox{ } 0.30)$, (4) $\mbox{RE}<\mbox{HY}\le\mbox{SW}<\mbox{SF}<\mbox{RA}$ in $[0.30 \mbox{ } 0.40)$, (5) $\mbox{RE}<\mbox{SW}\le\mbox{HY}\le\mbox{SF}<\mbox{RA}$ in $[0.40  \mbox{ } 0.62)$, and (6) $\mbox{RE}<\mbox{SW}<\mbox{SF}\le\mbox{HY}<\mbox{RA}$ in $[0.62  \mbox{ } 0.90]$ ($p<.001$).
In (f) $h_{vol}$, the order was $\mbox{RE} < \mbox{SW} < \mbox{RA} < \mbox{SF} \le \mbox{HY}$ for all sparisty  ($p<.001$).)  } \label{fig:binary_toy} \end{center} \end{figure}

Fig. \ref{fig:binary_toy} showed the results of (a) $h_{vol}$, (b) $h_{fun}$, (c) $Q$, (d) $h_{spe}$, (e) $e_{loc}$, and (f) $e_{glo}$. 
We plotted the box plots at the sparsity $0.09, 0.18, 0.27, 0.36, 0.45,$ and $0.54$ from left to right. 
In each figure, the horizontal and vertical axes represented the graph type and the network invariant, respectively.   
The color of line was changed by the type of graph: blue for RE, green for SW, red for RA, cyan for SF, and magenta for HY. 
In Fig. \ref{fig:binary_toy} (a) $e_{glo}$, the order of five graph types was changed four times at the sparsity  $0.04, 0.12, 0.20,$ and $0.48.$ The inconsistent order was due to SF and HY, and the order of RE, SW, and RA was comparatively consistent for the sparsity with RE $\le$ SW $\le$ RA. 
In (b) $e_{loc}$, the order of graph types was also changed four times at the sparsity $0.04, 0.08, 0.40, $ and $0.66$. $e_{loc}$ also consistently discriminated RE, SW, and RA in the order of RA $<$ SW $<$ RE.  
However, the order with SF and HY was inconsistent. 
In (c) $Q$, the order was $\mbox{RA} \le\mbox{SF}<\mbox{HY}<\mbox{SW}<\mbox{RE}$ for all sparsity ($p<.001$, FDR-corrected).
In (d), $h_{fun}$ of five network types was always the same at the fixed sparsity. 
It was because the functional entropy consider only the distribution of edge weights, not the network topology.  
In addition, the functional entropy did not monotonically increase or decrease over sparsity. 
Thus, it could also not distinguish the difference in the sparsity of graph.  
In (e) $h_{spe}$, the order of graph types in $h_{spe}$ was changed six times at the sparsity $0.04, 0.12, 0.16, 0.30, 0.40,$ and $0.62$. 
The order of graph types in $h_{spe}$ highly depended on the sparsity. 
$h_{spe}$ measured the connectedness of a graph. 
Because all nodes in RE had the same degree, RE did not have a modular structure and it always had the smallest $h_{spe}$ among all five graph types. 
In (f) $h_{vol}$, the order of graph types was consistent for all sparsity.  
The order was $\mbox{RE}<\mbox{SW}<\mbox{RA}<\mbox{SF}\le\mbox{HY}$ ($p<.001$, FDR-corrected). 
The volume entropy of SF was similar to that of HY at large sparsity. 
It was because that as the number of edges increased in a graph, SF lost its sparse property. 
The volume entropy also distinguished well between RE, SW, RA, SF, and HY.

\subsection{Using artificial weighted graphs} 
\label{sec:toy_weighted} 

In this simulation, we compared the six graph invariants in discriminating three distinct types of artificial weighted graphs. 
These three types of weighted graphs had the same topological structure, but they had different edge weights. 
We generated 150 hyperbolic unweighted graphs using CNM toolbox, and defined the edge distance of the graph in three different ways,  
\begin{itemize}[noitemsep,topsep=0pt]
\item Uniform edge distance (U): all edges had the same distance,   
\item Long edge distance with high node degrees (L): the edge distance was proportional to the degree of its initial and terminal nodes, $i(e)$ and $t(e)$, determined by   
\be
\label{eq:minhvol} 
l(e) = \frac{\log (k_{i(e)}-1) + \log(k_{t(e)}-1)}{\sum_{\forall v \in V} k_v \log (k_v-1)}, 
\ee 
where $k_v$ is the number of edges connecting with a node $v \in V$ \cite{lim.2008.tams}.  
\item Short edge distance with high node degrees (S): the edge distance was inversely proportional to the degree of two connected nodes, determined by the inverse of $l(e)$ in (\ref{eq:minhvol}). 
\end{itemize}
These three networks had the same topology, but different geometries.  
The edge connecting nodes with higher degree was longer in L, but shorter in S. 
Thus, it could be assumed that the information propagation was the fastest in S, followed by U and L. 
Note that before estimating the graph invariants, we normalized the volume of weighted graph to two, i.e., $\sum_{e} l(e) = 2$. 
 
\begin{figure}[t] \begin{center}
\includegraphics[width=1\linewidth]{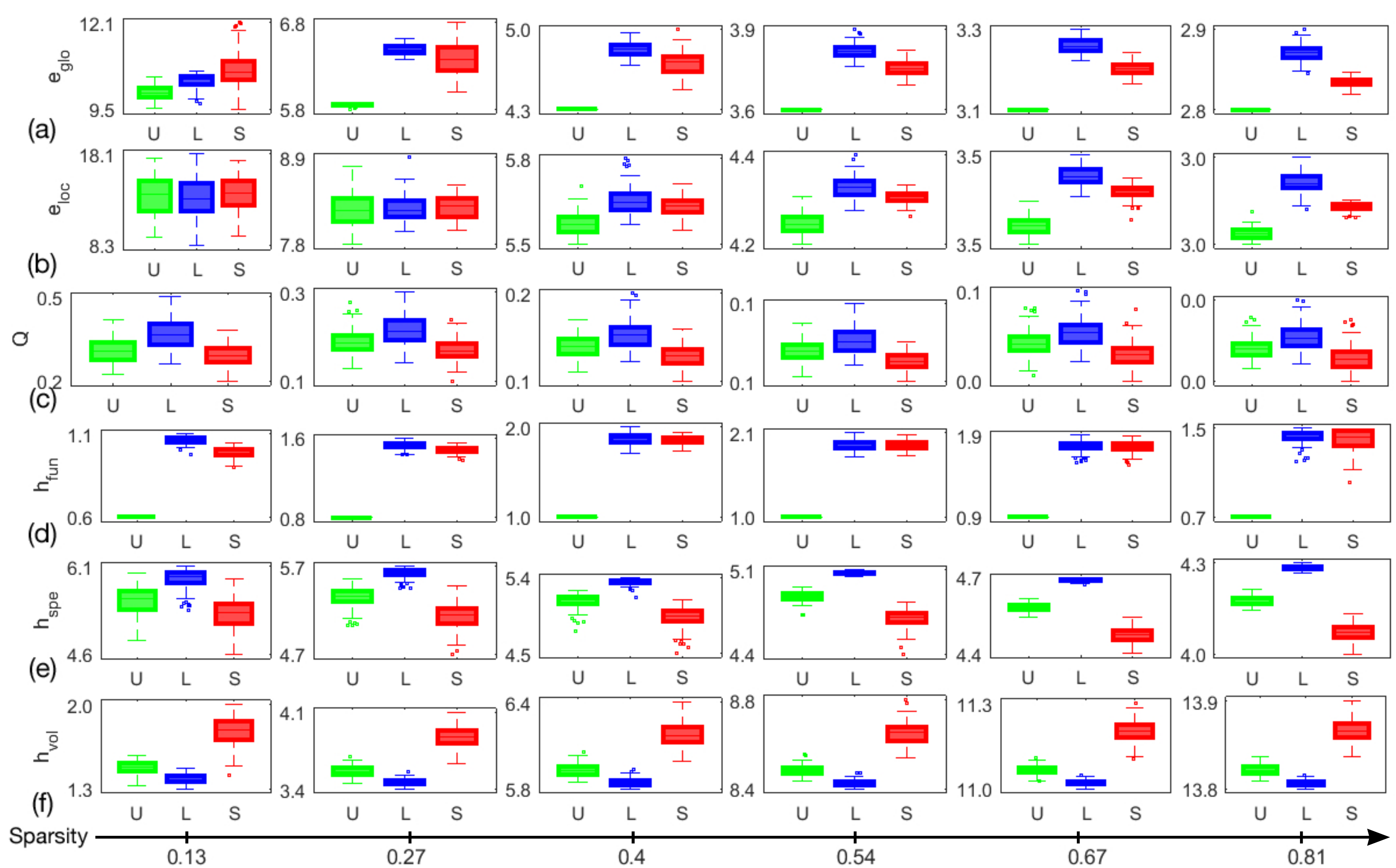} \caption{Comparison of graph invariants using various artificial weighted graphs. Each panel shows the box plot of (a) global efficiency ($e_{glo}$), (b) average local efficiency ($e_{loc}$), (c) modularity ($Q$), (d) functional entropy ($h_{fun}$), (e)  spectral entropy ($h_{spe}$), and (f) volume entropy ($h_{vol}$). The sparsity of weighted network was $0.13, 0.27, 0.40, 0.54, 0.67,$ and $0.81$ from left to right. The color of line represents U (green), L (blue), and S (red). The order of the three types of different weighted graphs was consistent for the sparsity when the graph property was measured by (c) $Q$, (e) $h_{spe}$, and (f) $h_{vol}$. However, the order was changed more than twice depending on sparsity in the other graph invariants, (a) $e_{glo}$, (b) $e_{loc}$, and (d) $h_{fun}$. \\ (In (a) $e_{glo}$, the order was (1) $\mbox{L}<\mbox{U}<\mbox{S}$ in the sparsity $[0.04  \mbox{ } 0.12)$, (2) $\mbox{U}<\mbox{L}<\mbox{S}$ in $[0.12  \mbox{ } 0.26)$, and (3) $\mbox{U}<\mbox{S}<\mbox{L}$ in $[0.26  \mbox{ } 0.90]$ ($p<.001$). 
In (b) $e_{loc}$, the order was (1) $\mbox{U} \approx \mbox{L} \approx  \mbox{S}$ in $[0.04  \mbox{ } 0.30)$, and (2) $\mbox{U}<\mbox{S}<\mbox{L}$ in $[0.30  \mbox{ } 0.90]$. 
In (c) $Q$, the order  was $\mbox{S}<\mbox{U}<\mbox{L}$ for all sparsity. 
In (d) $h_{fun}$, the order was  (1) $\mbox{U}<\mbox{S}<\mbox{L}$ in $[0.04  \mbox{ } 0.80)$, and (2) $\mbox{U}<\mbox{L}<\mbox{S}$ $[0.80  \mbox{ } 0.90]$ ($p<.001$). 
In (e) $h_{spe}$, the order was $\mbox{S}<\mbox{U}<\mbox{L}$  for all sparsity. 
In (f) $h_{vol}$, the order was $\mbox{L}<\mbox{U}<\mbox{S}$ for all sparisty  ($p<.001$).)  
}\label{fig:weighted_toy}  \end{center} \end{figure}

The results of graph invariants were shown in Fig. \ref{fig:weighted_toy}. 
In each figure, three weighted graphs, S, U, and L were represented by red, blue, and green, respectively. 
In (a) $e_{glo},$ the order of S, U, and L was changed three times at the sparsity $0.04, 0.12,$ and $0.26$. 
The order of graph types, S, U, and L was consistent for all sparsity in (b) $e_{loc}$, (c) $Q,$ (e) $h_{spe}$, and (f) $h_{vol}$. 
The order was $\mbox{L}<\mbox{U}<\mbox{S}$ in $h_{vol}$ in Fig. \ref{fig:weighted_toy} (f), $\mbox{S}<\mbox{U}<\mbox{L}$ in $Q$ and $h_{spe}$ in (c) and (e), and $\mbox{U} \le \mbox{S} \le \mbox{L}$ in $e_{loc}$ in (b) ($p<.001$, FDR-corrected). 
In (d), the order of graph types in $h_{fun}$ was changed twice at $0.04,$ and $0.80$. 
$h_{fun}$ could not find the difference between S and L.   


\subsection{Using artificial modular graphs} 
\label{sec:toy_modular} 

\begin{figure}[t] \begin{center}
\includegraphics[width=1\linewidth]{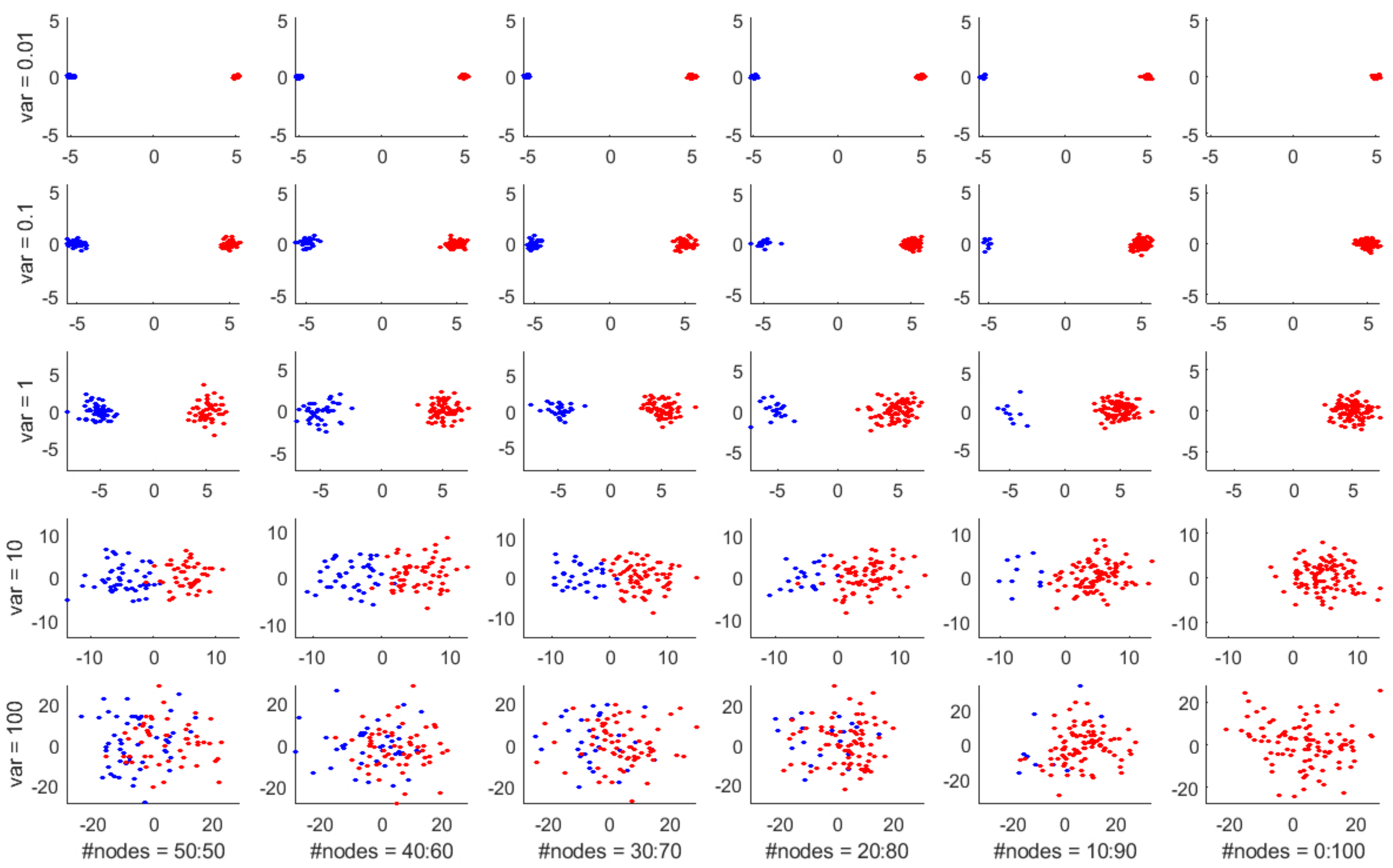} \caption{Example of artificial modular graphs. Two modules in a graph were generated by bivariate Gaussian distribution with mean $[-5, 0]$ and $[5,0]$. The variance was varied from $0.01$ to $100$ from top to bottom. The number of nodes in the graph was 100. The ratio of the number of nodes in two modules was varied from $50:50$ to $0:100$ from left to right. The color of nodes in two modules on the left and right sides was blue and red, respectively. 
}\label{fig:modular_toy}  \end{center} \end{figure}
\begin{figure}[t] \begin{center}
\includegraphics[width=1\linewidth]{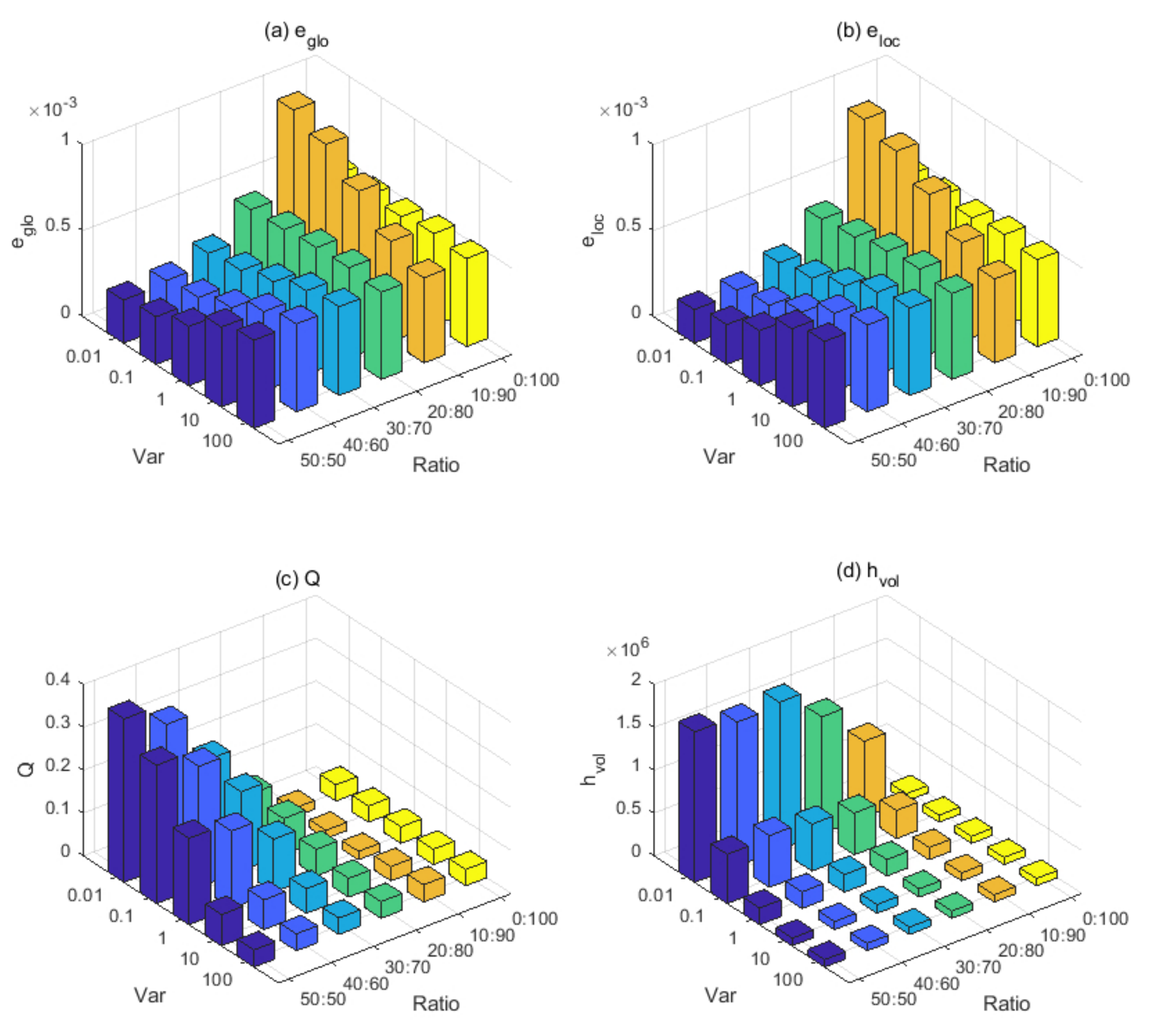} \caption{Average (a) $e_{glo}$, (b) $e_{loc}$, (c) Q, and (d) $h_{vol}$ of $100$ artificial modular graphs with respect to variance and ratio of the node numbers of two modules. The x- and y-axes represents the variance of bivariate Gaussian distribution that generated two modules and the ratio of the node numbers of two modules. Total number of nodes in a graph was $100$. 
}\label{fig:modular_invs}  \end{center} \end{figure}

In this simulation, we observed the volume entropy and the edge capacity by varying the modular structure of artificial graphs. 
We generated the modular graph with two modules which were generated by two bivariate Gaussian distributions with mean $[-5,0]$ and $[5,0]$, respectively. 
The variance of the distributions was varied by $0.01, 0.1, 1, 10,$ and $100$. 
The total number of nodes was $p=100$. 
The ratio of node numbers in two modules was changed by $50:50, 40:60, 30:70, 20:80, 10:90,$ and $0:100$. 
There was no module in a graph at the ratio $0:100$. 
The example of modular graphs was shown in Fig. \ref{fig:modular_toy}. 
In the figure, the ratio of node numbers was varied from left to right columns, and the variance was varied from top to bottom rows. 
In each panel, two modules had two different colors, blue on the left and red on the right. 
$100$ artificial modular graphs were generated at each variance and each ratio of node numbers. 
Then, the edge distance in a graph was estimated by Euclidean distance between any two nodes. 
After constructing 100 modular graphs at each variance and each ratio, the average $e_{glo}$, $e_{loc}$, $Q$, and $h_{vol}$ was estimated as shown in Fig. \ref{fig:modular_invs}. 
As the variance increased, $e_{glo}$ and $e_{loc}$ increased, and $Q$ decreased except for the ratio of node numbers $10:90$ and $0:100$. 
In constrast, $h_{vol}$ decreased as the variance increased at all ratios. 

\begin{figure}[t] \begin{center}
\includegraphics[width=1\linewidth]{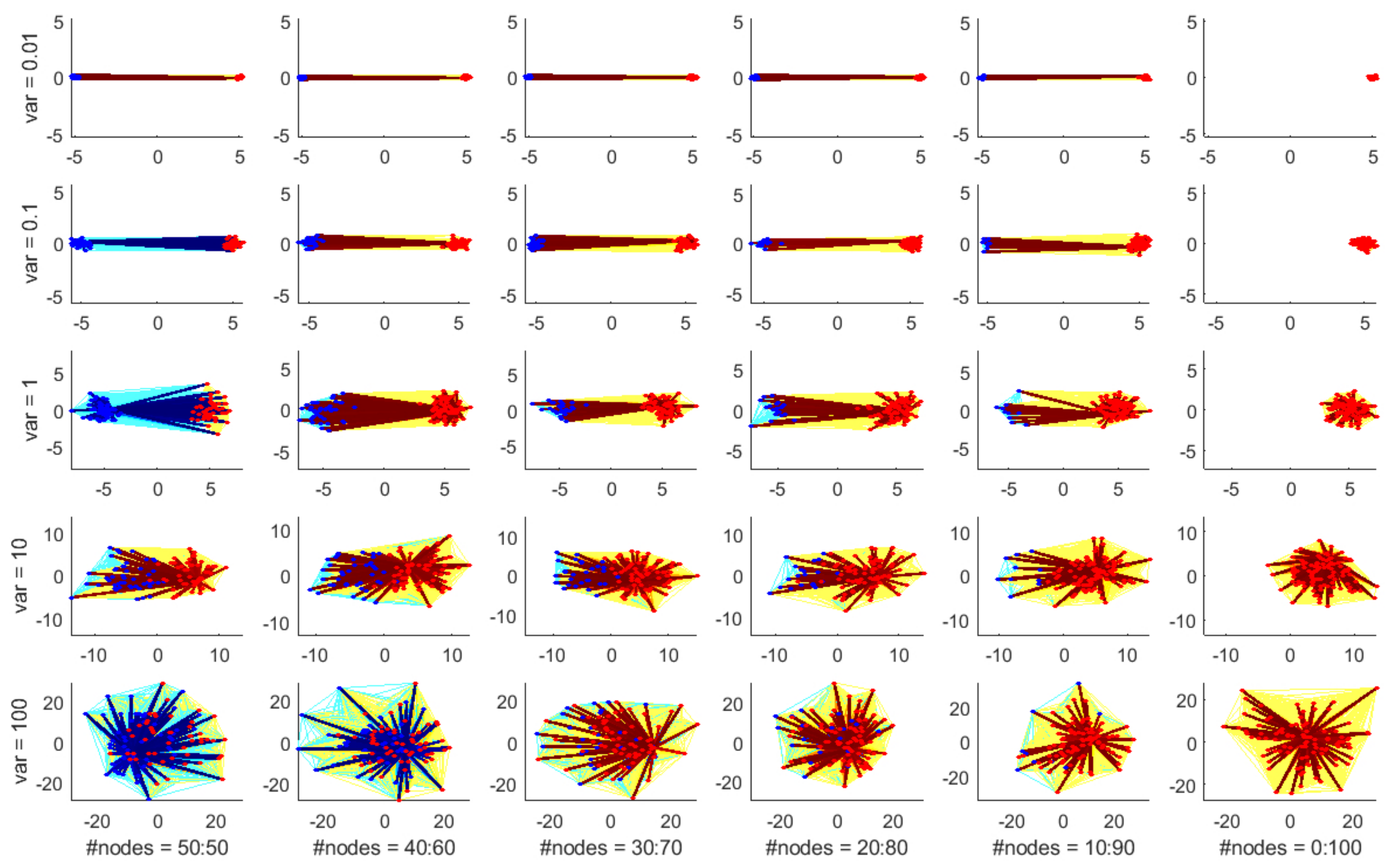} \caption{Directed graphs induced by the edge capacity matrix of artificial modular graphs in Fig. \ref{fig:modular_toy}. The color of an edge is the same as the color of the terminal node of the edge. The color of edge is changed from yellow to dark red or from cyan to dark blue as the edge capacity increases.
}\label{fig:modular_directed}  \end{center} \end{figure}

\begin{figure}[t] \begin{center}
\includegraphics[width=1\linewidth]{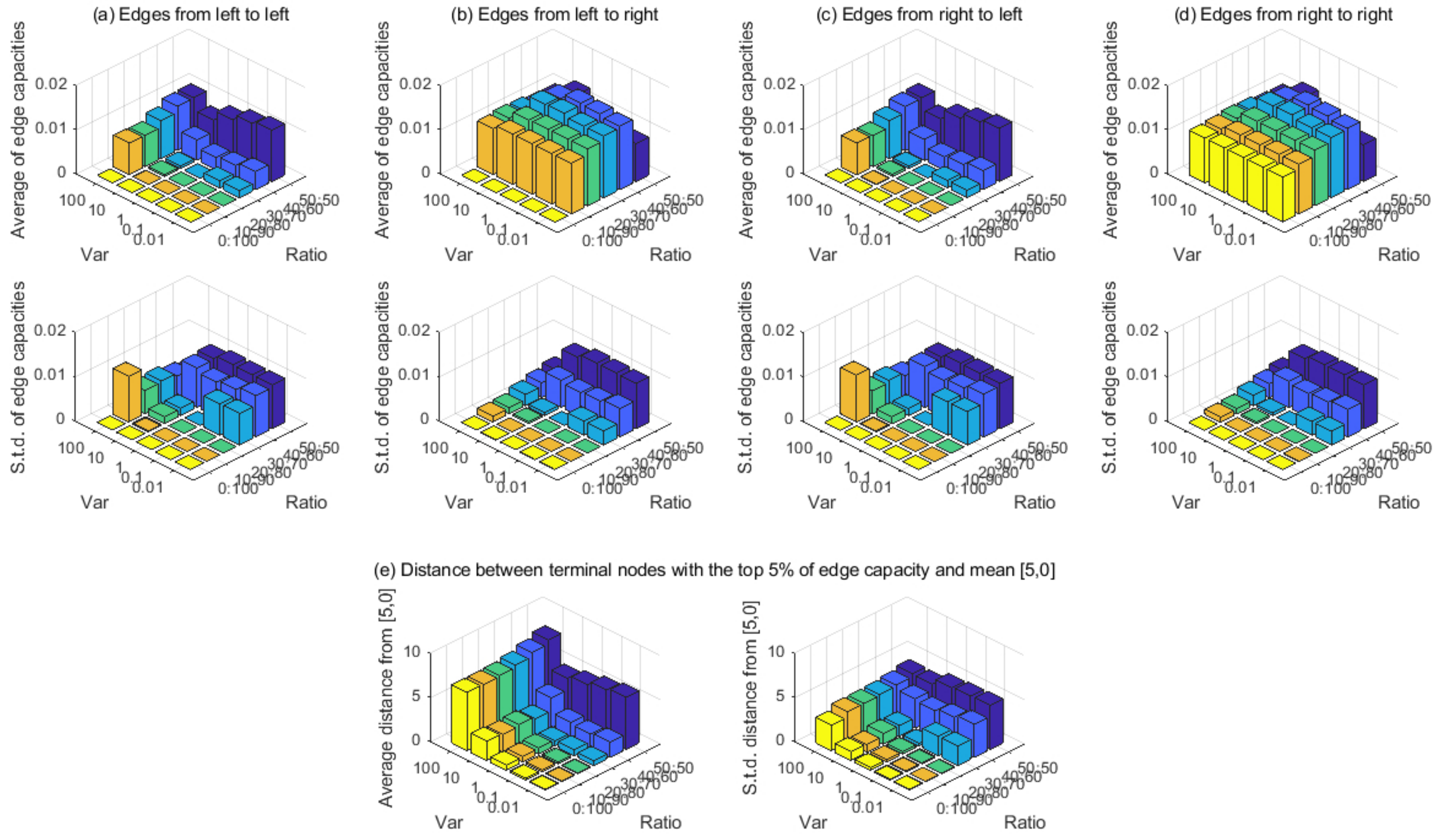} \caption{Average (upper) and standard deviation (lower) of edge capacities (a) from left to left modules, (b) from left to right modules, (c) from right to left modules, and (d) from right to right modules. (e) Average (left) and standard deviation (right) of the distance between the terminal node of the top 5 \% of edge capacity and the mean $[5,0]$ of the right module. 
}\label{fig:modular_directed_capacity}  \end{center} \end{figure}

We also estimated an edge capacity matrix for each modular graph. 
Fig. \ref{fig:modular_directed} showed the directed graph induced by the example of artificial modular graphs in Fig. \ref{fig:modular_toy}. 
Among bidirectional edges in the directed graph, the edge with larger capacity was plotted in each figure. 
The direction of edge was represented by the color of edge. 
If the edge was directed from the blue node on the left to the red node on the right, the color of edge was red. 
Otherwise, the color of edge was blue. 
As the edge capacity increased, the color of edge was changed from yellow to dark red or from cyan to dark blue. 
In Fig. \ref{fig:modular_directed}, as the number of nodes on the right module increased from left to right, the red edges from the left to the right modules were plotted. 
To compare the edge capacity between modules, we divided the edge capacity matrix into four block matrices, a sub-matrix for the edges from the left to the left modules, that from the left to the right modules, that from the right to the left modules, and that from the right to the right modules. 
The average and the standard deviation of each block matrix were plotted in Fig. \ref{fig:modular_directed_capacity} (a-d). 
The average edge capacities to the right modules in the upper part of (b) and (d) were larger than that to the left modules in the upper part of (a) and (c) except for the ratio $0:100$ and $50:50$ and the variance $100$. 
We also estimated the distance between the location of the terminal nodes with the top 5 \% edge capacities and the mean $[5,0]$ of the right module. 
We found that the terminal nodes with the top 5 \% edge capacities were close to the mean of the right module when the variance was in between $0.01$ and $1$ and the ratio was in between $0:100$ and $30:70$. 
The simulation results showed that more paths from the small-sized module to the large-sized module were generated than the opposite direction at the stationary state of a graph. 
Especially, the paths with the largest edge capacities were directed to the center of the large-sized module. 


\section{Results}

\subsection{Clinical dataset: resting state fMRI and PET}
\label{sec:clinical_hvol}

\begin{figure}[t] \begin{center}
\includegraphics[width=1\linewidth]{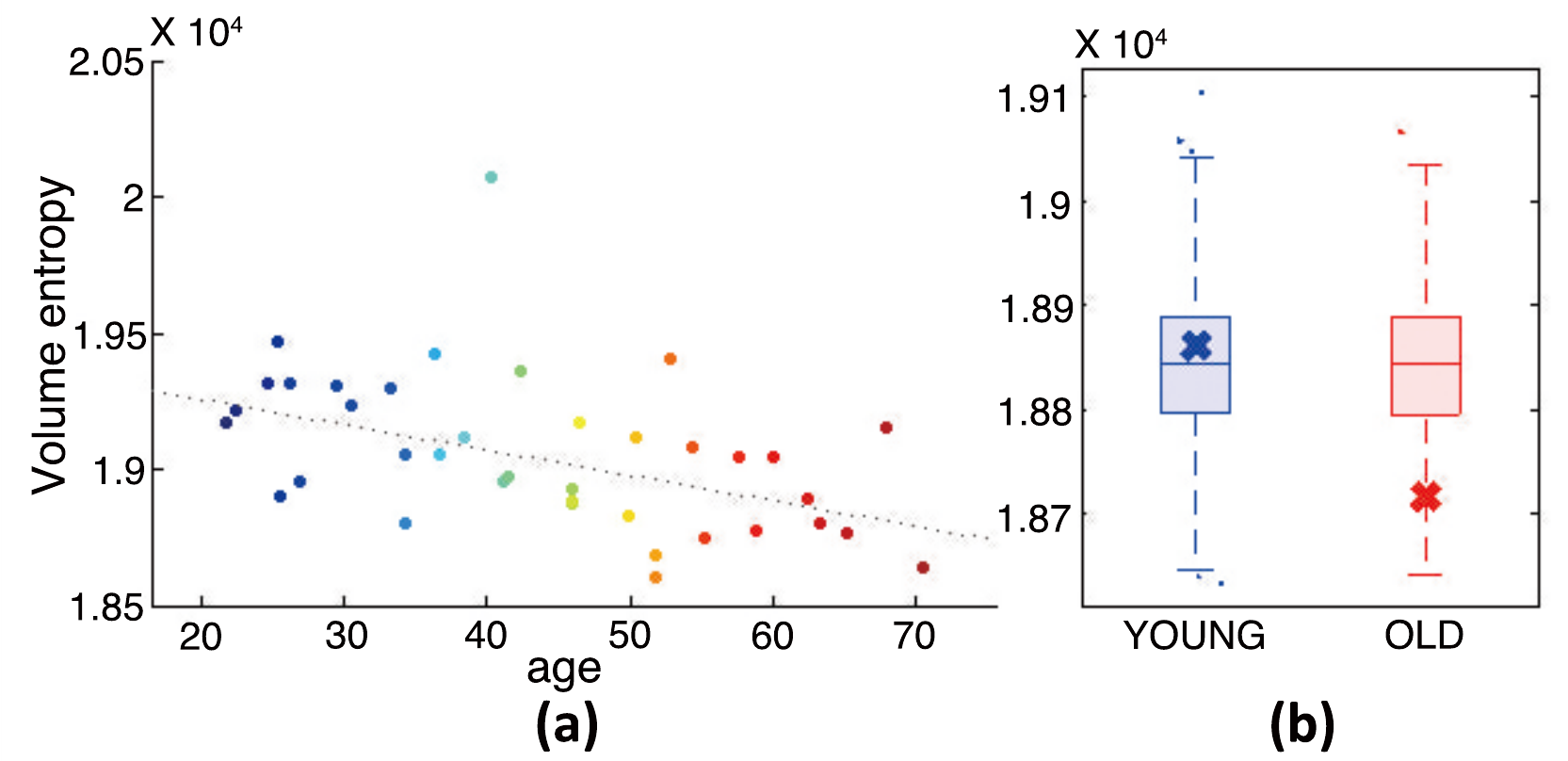} \caption{(a) Normalized volume entropy of 38 functional graphs of resting state fMRI with respect to age. The volume entropy was significantly correlated with age (negative correlation, $p<.005$). (b) Normalized volume entropy of Y and O in the metabolic graphs of PET. Box plots showed the volume entropies of 5000 null graphs constructed by permuted PET data set. The green marker 'X' represented the volume entropy of true metabolic graphs, Y and O. The volume entropy of O was significantly different from that of null graphs, but the volume entropy of Y was not ($p<.05$). The difference of the volume entropy between Y and O was not significant, but showed the tendency of Y $>$ O ($p<.13$). } \label{fig:aging_measures} \end{center} \end{figure}

The normalized volume entropy of 38 functional graphs was plotted with respect to age in Fig. \ref{fig:aging_measures} (a). 
The volume entropy and the age were negatively correlated ($p<.005$). 
The normalized volume entropy of Y and O in metabolic graphs was shown in Fig. \ref{fig:aging_measures} (b) by green marker `X'. 
In the metabolic graph analysis, we performed 5000 permutations of Y and O to enable the assessment of statistical differences between the two groups. 
If we called a graph constructed by permutation a null graph, the box plot in Fig. \ref{fig:aging_measures} (b) showed the volume entropies of 5000 null graphs. 
The volume entropy of O was significantly smaller than that of null graphs, but the volume entropy of Y was not ($p<.05$).  
The difference between Y and O was not significant, but showed the tendency that the volume entropy of Y was larger than that of O ($p<.13$). 
The results of both functional and metabolic graphs showed that the volume entropy decreased with normal aging. 

\subsection{Edge capacity matrix on a metabolic graph}
\label{sec:flow_pet} 

\begin{figure*}[t] \begin{center}
\includegraphics[width=1\linewidth]{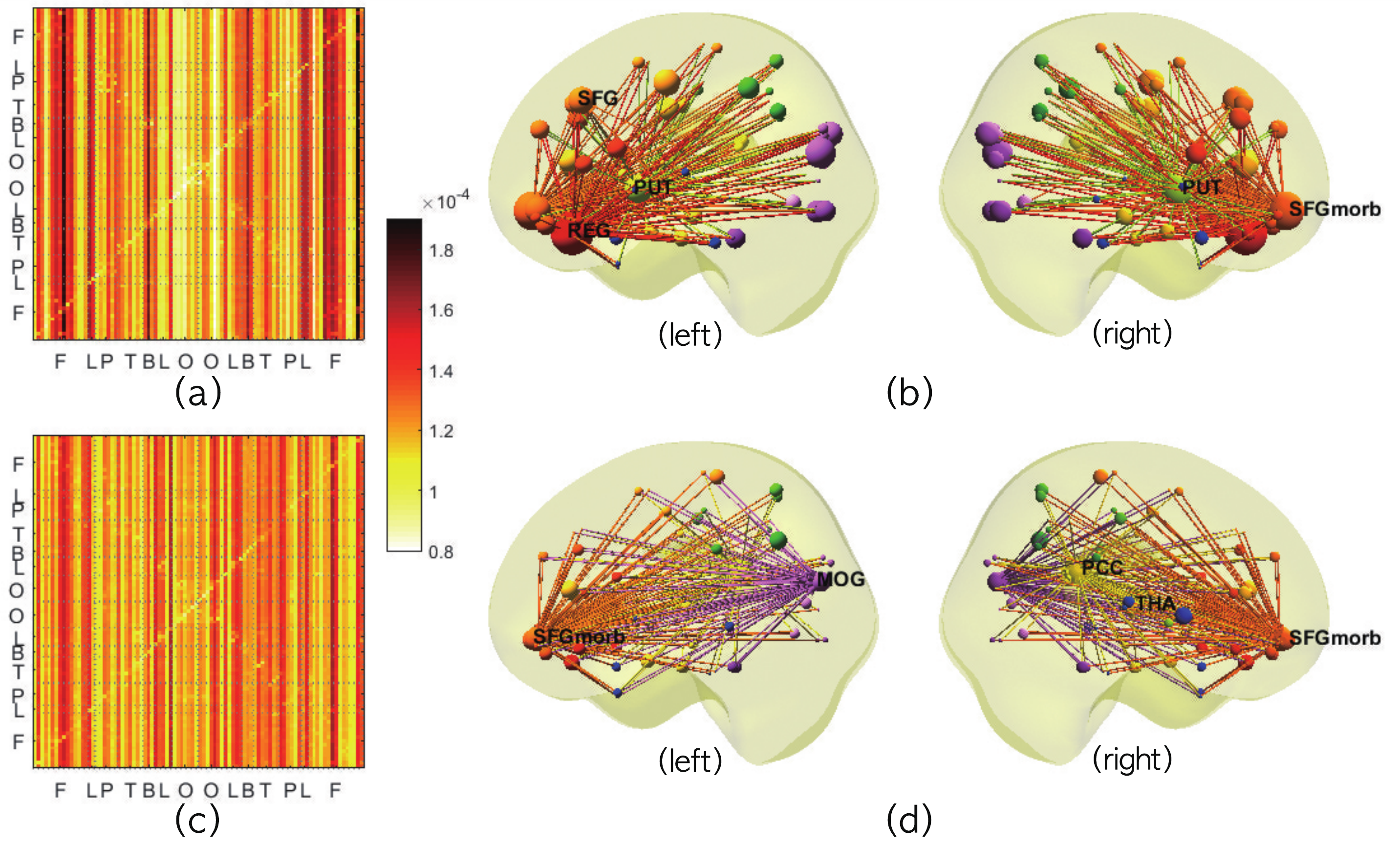} \caption{(a,c) Edge capacity matrices of Y and O in PET. In the edge capacity matrix, the first 45 rows and columns corresponded to right hemisphere and the last 45 rows and columns corresponded to left hemisphere. F, L, P, T, B, L, and O represented frontal, limbic (cingulate cortex), parietal, temporal, basal ganglia, limbic (hippocampus and parahippocampal gyrus), and occipital lobes. (b,d) Directed graphs of Y and O. Only the top 5 \% of edge capacity were plotted in the directed graph. The color of node represents the location of node (more details in the supplementary material). The size of node was proportional to the absolute value of node capacity. The color of edge was determined by the color of its terminal node. } \label{fig:flow_network_pet} \end{center} \end{figure*}

Fig. \ref{fig:flow_network_pet} showed the edge capacity matrix and the directed graphs of the metabolic graph of Y and O. 
The edge capacity matrices of Y and O were shown in Fig. \ref{fig:flow_network_pet} (a) and (c), respectively. 
The obtained directed graphs of Y and O were shown in (b) and (d), respectively.    
In the edge capacity matrix,  the first 45 rows and columns were the nodes in right hemisphere, and the last 45 rows and columns were in left hemisphere.   
The nodes were sorted in the order of the frontal (F), limbic (L), parietal (P), temporal (T), basal ganglia (B), limbic (L), and occipital (O) lobes (more details in the supplementary material).  
The $(i,t)$th entry of the edge capacity matrix was the edge capacity directed from the node $i$ to $t$. 
As the edge capacity decreased, the color of entry was changed from dark red through yellow to white as shown in the right colorbar. 
In the edge capacity matrices in Fig. \ref{fig:flow_network_pet} (a) and (c), each column had similar color. 
It meant that the edges connected to the same terminal node had similar edge capacity. 

In Fig. \ref{fig:flow_network_pet} (b) and (d), we plotted only edges with the top 5 \% of edge capacity in the directed graphs of Y and O.   
In each figure, the left and right panels showed the same brain graph in the left and right views, respectively. 
In the directed graph of Y in (b), the edges were mainly directed to the medial orbital part of superior frontal gyrus (SFGmorb) in the right hemisphere, bilateral putamen (PUT), left dorsolateral superior frontal gyrus (SFG), and left gyrus rectus (REG). 
In the directed graph of O in (d), the edges were mainly directed to bilateral SFGmorb, right thalamus (THA), right posterior cingulate cortex (PCC), and left middle occipital gyrus (MOG). 
The color of node represented the location of node: red and orange in F, green in P, blue in T, purple in O, yellow in L, and yellow-green in B (more details in the supplementary material). 
The size of node was determined by the absolute value of node capacity.  
The color of edge was the same as the color of the terminal node of the edge. 


We performed 5000 permutations and Wilcoxon rank sum test for finding the difference between the edge capacities of Y and O.  
There was no significant edges that had larger capacity in Y than in O. 
In contrast, the information capacity of Y$<$O were found in the connections from the most of brain regions to left angular gyrus (ANG) ($p<.05$, FDR-corrected).
The node capacity of left ANG was also larger in O than in Y ($p<.05$, FDR-corrected). 

\subsection{Edge capacity matrix on a functional graph}
\label{sec:flow_fmri} 

\begin{figure*}[t] \begin{center}
\includegraphics[width=1\linewidth]{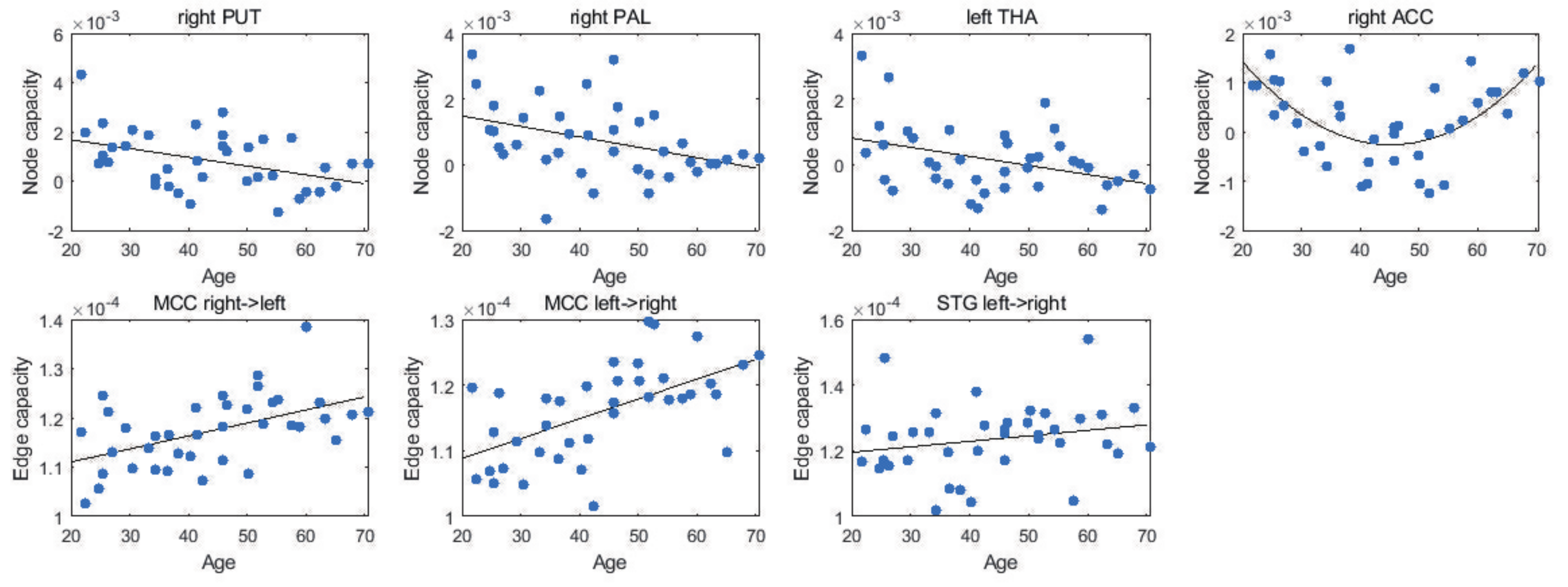} \caption{(Upper) Node capacity of right PUT and PAL, left THA, and right ACC with respect to age in 38 functional graphs. The node capacity of right PUT and PAL, and left THA significantly decreased with age, and that of right ACC was a U-shaped curve with age ($p<.05$, FDR-corrected).  (Lower) Edge capacity directed from right MCC to left MCC, in the opposite direction, and from left STG to right STG with respect to age. The edge capacity increased with age ($p<.05$, FDR-corrected).} \label{fig:node_cap_fmri} \end{center} \end{figure*}

The directed graphs of 38 subjects in the resting state fMRI were shown in the supplementary material. 
We estimated edge capacities that were significantly correlated with age.  
The negative correlation with the age were found in the edges directed from the most of brain regions to right PUT and pallidum (PAL), and left THA ($p<.05$, FDR-corrected). 
The node capacity of right PUT and PAL, and left THA also decreased with age  as shown in Fig. \ref{fig:node_cap_fmri} ($p<.05$, FDR-corrected). 
The edge capacities to left PUT and PAL, and right THA and their node capacities also tended to be negatively correlated with age ($p<.05$, uncorrected).  

The positive correlation with age were found in the bidirectional edges between left and right median cingulate cortex (MCC) and the edge from left superior temporal gyrus (STG) to right STG as shown in Fig. \ref{fig:node_cap_fmri} ($p<.05$, FDR-corrected). 
We also estimated a quadratic relationship between the edge capacity and the age. 
The capacity of the most of edges directed to right anterior cingulate cortex (ACC) had a U-shaped curve with respect to age ($p<.05$, FDR-corrected). 
It decreased to around 45 years of age and increased at older age.   
The node capacity of right ACC also had a U-shaped curve with respect to age. 
The minimum node capacity of right ACC was also found at around 45 years of age as shown in Fig. \ref{fig:node_cap_fmri} ($p<.05$, FDR-corrected). 


\subsection{Global and local efficiencies, modularity, and age} 
\label{sec:existing_measure}


To see the relationship between the volume entropy and the existing complex graph invariants, we also estimated the global and local efficiencies, and modularity. 
The global and average local efficiency highly depended on the volume of graphs. 
When we estimated the unnormalized global and average local efficiency, both of them significantly increased with age in the resting state fMRI ($p<.05$). 
However, when we estimated them after the normalization of graph volume, the normalized global efficiency tended to decrease with age ($p = 0.085$), but the normlaized local efficiency tended to increase with age ($p=0.070$). 
The volume of 38 functional graphs decreased with age ($p<.05$). 
The unnormalized volume entropy of 38 functional graphs had no relationship with age. 
The modularity decreased with age in the resting state fMRI regardless of the normalization of graph volume ($p<.05$). 
There was no node that was significantly related to age in node strength and local efficiency in functional graphs. 
In the metabolic graph of Y and O, there was no difference in the volume, global and local efficiencies, and modularity.  
 

\section{Discussion} 

\subsection{Relationship between volume entropy and complex graph measures} 
\label{sec:relationship_networkmeasures}

To better understand the volume entropy, we discuss the relationship between the volume entropy and the existing complex graph invariants such as modularity, global and local efficiencies, and hubs. 
Firstly, the volume entropy was large when there were many edges in a graph. 
The simulation in Sec. \ref{sec:toy_binary} and \ref{sec:toy_weighted} showed that the change of sparsity affected the volume entropy more than the change of network topology and geometry.   
These results were found not only in the volume entropy but also in the other graph invariants such as global and local efficiencies and spectral and functional entropies. 
Because we used fully connected weighted graphs in the brain imaging data applications, there was no effect on the volume entropy from the difference of sparsity. 

Secondly, the volume entropy was more related to global efficiency than local efficiency. 
When the volume entropy was applied to the graphs with distinct topology in Sec. \ref{sec:toy_binary}, the order of graphs was $\mbox{RE}<\mbox{SW}<\mbox{RA}$ for all sparsities.  
The global efficiency was proportional to the characteristic path length, while the local efficiency was inversely proportional to the clustering coefficient \cite{rubinov.2009.ni}. 
According to the Watts-Strogatz model of the small world, the characteristic path length and the average clustering coefficient were the smallest in RE, followed by SW and RA \cite{watts.1998.nature}. 
Our results in Sec. \ref{sec:toy_binary} also showed that the global efficiency was the smallest in RE, followed by SW and RA, while the local efficiency was the opposite. 
If a graph had high average clustering coefficient, but short characteristic path length, the information would not be propagated throughout the graph because the information would whirl around only in the nodes with strong clustering coefficients. 
That might be the reason why the volume entropy of SW was smaller than that of RA. 

Thirdly, the volume entropy was large when a graph had hubs. 
SF and HY had larger volume entropy than RE, SW, and RA in Sec. \ref{sec:toy_binary}. 
SF and HY were a graph with hubs that played a decisive role in the exponential growth of the path in a graph through which information was delivered \cite{eguiluz.2005.phrl,krioukov.2010.pre}. 
Fourthly, if the graphs had similar global efficiencies, the volume entropy could vary depending on the local efficiencies of the graphs. 
The volume entropy of HY was larger than that of SF in Sec. \ref{sec:toy_binary}. 
HY was known as a network with high clustering coefficient and heterogeneous degree distribution, while SF had only heterogeneous degree distribution \cite{krioukov.2010.pre}.  
There were many paths between highly clustered nodes. 
If the paths outgoing from the clustered nodes were appropriately created, the high local efficiency could also contribute to fast information propagation.  
In this sense, the volume entropy may be the first global invariant to measure the efficiency of hyperbolic graph. 
 
Finally, the volume entropy was also related to the modular structure of network as shown in Sec. \ref{sec:toy_modular}. 
Nodes within a module were densely connected and the shortest path length between any nodes within a module was short. 
In contrast, nodes between modules were loosely connected and the shortest path length between nodes in different modules was long. 
Since the clustering coefficient and the characteristic path length were estimated in an average manner, they were not proper to represent heterogeneous shortest path length and heterogeneous connected nodes in a modular graph.  
However, the volume entropy was calculated by the fastest growth rate of paths in a graph, and not affected by such a heterogeneous property of a graph.

\subsection{Normalization of graph volume}

The normalized volume entropy significantly decreased with age. 
However, 
the unnormalized volume entropy had no relationship with age because the volume of functional graphs decreased with age in Sec. \ref{sec:existing_measure}. 
The decline of brain graph volume with age might mean that the connection between any brain regions was generally shorter. 
The decline of the normalized volume entropy might mean that the inherent topological structure of the brain graph became increasingly inefficient. 
Since the unnormalized volume entropy had no relationship with age, 
it could be interpreted that the connections in the functional brain graph became shorter, i.e., the correlations between brain regions became stronger with age in order to compensate the inefficient topological change of brain graph across the lifespan.

\subsection{Comparison of the results with the previous studies}

Previous studies on resting-state functional connectivity have shown somewhat inconsistent change of global and local efficiencies across the lifespan \cite{cao.2014.dcn,ferreira.2013.nbr,sala-llonch.2015.fp,zuo.2017.tcs}. 
The human brain has known to have a modular architecture \cite{chen.2008.cc,sporns.2016.arp,valencia.2009.chaos}. 
As discussed in Sec. \ref{sec:relationship_networkmeasures}, the modular network tended to have heterogeneous shortest path lengths and heterogeneous connected nodes. 
Thus, the global and average local efficiencies of modular brain graph have not been proper to measure the property of modular brain graph. 

In contrast, there were consistent reports of the age-related reorganization in the modular structure of functional connectivity \cite{cao.2014.dcn,geerligs.2015.cc,song.2014.bc}.
Especially, they have consistently shown that the modularity decreased after 40 years of age \cite{cao.2014.dcn,geerligs.2015.cc,song.2014.bc}. 
The results of our resting state fMRI data also showed the age-related decline of modularity in Sec. \ref{sec:existing_measure}. 
The age-related change in modularity 
might be related to the age-related inefficient topological change, which was also well-quantified by the volume entropy. 


\subsection{Edge capacity on a metabolic graph} 


The sum of edge capacity in a brain graph is one because it is the stationary distribution of the generalized Markov system in (\ref{eq:eig_stationary}). 
Therefore, the increase or decrease of the edge capacity with age should be interpreted as the change of the relative proportion of the edge capacity in the whole brain, not the change of its absolute value. 
We assumed that the information flowed through the paths in the graph, and the amount of information going through the edge was proportional to the number of paths on the edge. 

The result in Sec. \ref{sec:flow_pet} showed that the role of left ANG became more important in the information propagation with age in a metabolic graph. 
The large-sized module had large information capacity because it had more paths. 
The edge capacity from the small-sized module to the large-sized modules was much larger than that with the opposite direction as shown in Sec. \ref{sec:toy_modular}. 
Especially, the edges directed to nodes at the center of the large-sized module had larger edge capacity. 
Thus, it could be assumed that the size of the module including left ANG was larger in O than in Y, and the amount of information coming into the module of left ANG would also increase. 
In addition, more information would flow into ANG which was known as the functional hub of DMN \cite{andrews-hanna.2014.anyas}. 
The reason why only the left ANG had large information capacity might be because the left hemisphere had less age-related decline than the right hemisphere \cite{dolcos.2002.nbr}. 


\subsection{Edge capacity on a functional graph} 

The functional graph had a topological structure where the information propagation slowed down along with age. 
At the same time, the contributions of PUT, PAL, and THA to information propagation decreased with age. 
Previous studies consistently indicated that the circuit linking PUT, PAL, THA, and cortical areas played a key role in motor ability across the human lifespan \cite{alexander.1986.arn,manza.2015.ni}. 
The functional and structural alterations in the basal ganglia-thalamocortical circuits have been found in the progression of Alzheimer's disease and Parkinson's disease as well as normal aging \cite{coxon.2010.cc,dejong.2008.brain,garg.2015.fn,manza.2015.ni}. 

ACC has been known as a key area involved in cognitive and emotional processing \cite{bush.2010.tcs,cao.2014.fan}. 
Previous study on the resting-state fMRI showed that the decreased functional connectivity between ACC and default mode network would be associated with the deficit of cognitive processing in aging, while the increased functional connectivity between ACC and the emotion-related brain regions such as STG, inferior frontal gyrus (IFG), PUT, and amygdala (AMYG) would be associated with the well-maintained emotional well-being in aging \cite{cao.2014.fan}. 
In contrast, our result showed that the role of right ACC in information propagation decreased until around 45 years of age, but increased at the older age.  
This result was somewhat different from the previous studies, and its biological meaning needs further discussion in the future. 

The information capacities of bidirectional edges between right and left MCCs had a linear relationship with age. 
In our results, the edge capacity between bilateral brain regions tended to be slightly smaller than the other edge capacities. 
This might be because the bilateral brain regions were highly correlated and likely to be in the same module.  
If the node capacity of the bilateral MCCs had significantly increased with age, it could be interpreted that the role of the bilateral MCCs became increasingly important with age. However, since only the edge capacity between bilateral MCCs increased with age, we assumed that the bilateral MCCs consistently belonged to a module, and that the role of the module itself became increasingly important.
MCC has been known to be related to environmental monitoring and response selection \cite{angela.2007.sn,apps.2013.fn}. 
Therefore, it could be speculated that there was the age-related change in social decision-making of human \cite{lim.2015.fan}.

The edge capacity from left to right STGs also increased with age. 
In the brain imaging data applications, the node capacity of only right STG tended to increase with age, while that of left STG was not changed  ($p<.05$, uncorrected). 
From the result, we inferred that while the contribution of right STG slightly increased, but left STG did not.  
STG has been known to be involved in language processing, multisensory integration, and social perception \cite{hein.2008.jcn,jou.2010.br}. 
Especially, the dysfunction of right STG has been found to be related with the social cognition deficit in normal aging \cite{moran.2012.jn}.

\subsection{Limitations and conclusions} 

In our study, we introduced a new network invariant, called a volume entropy. 
It measured the fastest growth rate of paths in a graph through which the information was propagated over a brain.  
The larger the volume entropy was, the more information was propagated in a graph. 
Thus, it could be regarded as a new graph invariant of efficiency in terms of the information propagation. 
The simulation results showed that the volume entropy was proper to measure the efficiency of a graph with heterogeneous property such as modular and hyperbolic graphs. 
The information flow in a graph was modelled by the generalized Markov system associated with a newly defined edge-transition matrix. 
The volume entropy was estimated by the stationary equation of the generalized Markov system. 
At the same time, we could obtain the stationary distribution of information flow in a graph. 
It provided a new insight of how much and in what direction the information flowed on a brain. 

However, the edge capacity highly depended on the terminal node of the edge. 
Thus, the node capacity, which was the difference between the nodes' incoming and outgoing edge capacities, was sometimes enough to represent the directed graph induced by the stationary distribution of the generalized Markov system. 
In addition, the direction and capacity of edge were difficult to interpret their biological meaning.  
If we mathematically prove the relationship between the edge capacity and the existing complex network measures, it can be easier to interpret its biological meaning. 
In the results, the significance in the difference between Y and O in metabolic graphs was rarely found due to the small number of subjects. 
The tendency of the volume entropy to decrease with age was similar for both two modalities, however, the local changes in the directed graphs of the functional and metabolic graphs were quite different. 
We will improve the proposed method to enable multi-modal graph analysis to exploit the advantage of simultaneously acquired PET and fMRI data in the future. 
The proposed method can be applied to the brain imaging data of normal control as well as to that of various disease groups. 
We also expect that the proposed method can reveal the information flow of the effective functional connectivity of which connection represents the causal relationship between brain regions. 


\section{Acknowledgement} 

This work is supported by Basic Science Research Program through the National Research Foundation (NRF) (No.2013R1A1A2064593 and No.2016R1D1A1B03935463), NRF Grant funded by MSIP of Korea (No.2015M3C7A1028926 and No.2017M3C7A1048079), and NRF grant funded by the Korean Government (No. 2016R1D1A1A02937497, No.2017R1A5A1015626, and No.2011-0030815). 

\section*{References}

\bibliography{leehk}

\end{document}